 \documentclass[aps,prl,twocolumn,superscriptaddress]{revtex4-1}
\usepackage{graphicx}
\usepackage{color}
\usepackage{amsfonts}
\usepackage{isomath}
\usepackage{amsmath}
\usepackage{amsthm}
\usepackage{amsfonts}
\usepackage{braket}
\usepackage{yhmath}
\usepackage{bm}

\renewcommand{\vec}[1]{\mathbfit{#1}}

\def\kbar{$\bar{\mathrm{K}}$}

\def\kbarp{$\bar{\mathrm{K}}^{\prime}$}
\def\gbar{$\bar{\mathrm{\Gamma}}$}
\def\kgk{\kbar-\gbar-\kbarp}
\def\WS2{WS$_2$}
\def\MoS2{MoS$_2$}{

\begin{document}

\title{Momentum-resolved linear dichroism in bilayer MoS$_2$}

\author{Klara Volckaert}
\thanks{These authors contributed equally to the work.}  
\affiliation{Department of Physics and Astronomy, Interdisciplinary Nanoscience Center, Aarhus University,
8000 Aarhus C, Denmark}
\author{Habib Rostami}
\thanks{These authors contributed equally to the work.}  
\affiliation{Nordita, Center for Quantum Materials, KTH Royal Institute of Technology and Stockholm University, Roslagstullsbacken 23, SE-106 91 Stockholm, Sweden}
\author{Deepnarayan Biswas}
\affiliation{SUPA, School of Physics and Astronomy, University of St Andrews,
St Andrews KY16 9SS, United Kingdom}
\author{Igor Markovi\'c}
\affiliation{SUPA, School of Physics and Astronomy, University of St Andrews,
St Andrews KY16 9SS, United Kingdom}
\affiliation{Max Planck Institute for Chemical Physics of Solids, N\"othnitzer Str. 40, 01187 Dresden, Germany}
\author{Federico Andreatta}
\affiliation{Department of Physics and Astronomy, Interdisciplinary Nanoscience Center, Aarhus University,
8000 Aarhus C, Denmark}
\author{Charlotte E. Sanders}
\author{Paulina Majchrzak}
\author{Cephise Cacho}
\author{Richard T. Chapman}
\author{Adam Wyatt}
\author{Emma Springate}
\affiliation{Central Laser Facility, STFC Rutherford Appleton Laboratory, Harwell 0X11 0QX, United Kingdom}
\author{Daniel Lizzit}
\author{Luca Bignardi}
\thanks{current address: Department of Physics, University of Trieste, Via Valerio 2,Trieste 34127, Italy}
\author{Silvano Lizzit}
\affiliation{Elettra-Sincrotrone Trieste, S.S. 14 Km 163.5, Trieste 34149, Italy}
\author{Sanjoy K. Mahatha}
\author{Marco Bianchi}
\author{Nicola Lanata}
\affiliation{Department of Physics and Astronomy, Interdisciplinary Nanoscience Center, Aarhus University,
8000 Aarhus C, Denmark}
\author{Phil D. C. King}
\affiliation{SUPA, School of Physics and Astronomy, University of St Andrews,
St Andrews KY16 9SS, United Kingdom}
\author{Jill A. Miwa}
\affiliation{Department of Physics and Astronomy, Interdisciplinary Nanoscience Center, Aarhus University,
8000 Aarhus C, Denmark}
\author{Alexander V. Balatsky}
\affiliation{Nordita, Center for Quantum Materials, KTH Royal Institute of Technology and Stockholm University, Roslagstullsbacken 23, SE-106 91 Stockholm, Sweden}
\author{Philip~Hofmann}
\affiliation{Department of Physics and Astronomy, Interdisciplinary Nanoscience Center, Aarhus University,
8000 Aarhus C, Denmark}
\author{S{\o}ren~Ulstrup}
\email[Electronic address: ]{ulstrup@phys.au.dk}  
\affiliation{Department of Physics and Astronomy, Interdisciplinary Nanoscience Center, Aarhus University,
8000 Aarhus C, Denmark}

\newpage

\begin{abstract} Inversion-symmetric crystals are optically isotropic and thus naively not expected to show dichroism effects in optical absorption and photoemission processes. Here, we find a strong linear dichroism effect (up to 42.4\%) in the conduction band of inversion-symmetric bilayer MoS$_2$, when measuring energy- and momentum-resolved snapshots of excited electrons by time- and angle-resolved photoemission spectroscopy. We model the polarization-dependent photoemission intensity in the transiently-populated conduction band using the semiconductor Bloch equations and show that the observed dichroism emerges from intralayer single-particle effects within the isotropic part of the dispersion. This leads to optical excitations with an anisotropic momentum-dependence in an otherwise inversion symmetric material.  
\end{abstract}

\maketitle
Optical selection rules in absoprtion experiments are powerful tools that can be used to determine the symmetry of electronic states in solids \cite{Henderson:2006aa}. Given the similarity of the processes underlying optical absorption and photoemission, selection rules have also been exploited in angle-resolved photoemission spectroscopy (ARPES) for decades \cite{Eberhardt:1980aa}. More recently, optical dichroism has been used in ARPES to study orbital degrees of freedom \cite{Gierz:2012,Zhu:2013aa,Cao:2013}, as well as the Berry curvature of the initial Bloch  states \cite{Schuler:2019,Cho:2018}. A particularly interesting opportunity for polarization-dependent excitations arises in single-layer (SL) transition metal dichalcogenides (TMDCs) such as MoS$_2$, where the helicity of circularly polarized light strongly couples to the valley and spin degrees of freedom \cite{Xiao:2012ab,Mak:2014aa}, permitting the generation of a finite valley polarization \cite{makcontrol2012,zengvalley2012}.


Adding time-resolution (TR) to ARPES in a pump-probe experiment leads to a process involving two optical excitations. This opens the possibility of exploiting not only the selection rules governing the photoemission process, but also those giving rise to the initial optical excitation into a transiently populated conduction band (CB) state \cite{Perfetti:2007,Malic:2011,Rohwer:2011,johannsen:2013,Antonija-Grubisic-Cabo:2015aa,Bertoni:2016,Aeschlimann:2017}. Indeed, the creation of a finite valley-polarization in SL WS$_2$ has recently been followed in momentum space by TR-ARPES using circularly polarized pump pulses \cite{Ulstrup2017,Beyer:2019aa}. Here, we extend such an experiment to the case of bilayer (BL) MoS$_2$ in the 2H structure, which is inversion-symmetric and hence an optically isotropic material. Surprisingly, we observe a substantial linear dichroism effect. Our results can be reconciled with the momentum-dependent excited state population determined by solving the semiconductor Bloch equations within the framework of a low energy $\vec{k} \cdot \vec{p}$ model that accounts for intra- and interlayer interactions between the $d$ orbitals forming the valence band maximum (VBM) and conduction band minimum (CBM) around \kbar~(\kbarp). These findings underline the necessity of accounting for selection rules governing the optical excitation in addition to the photoemission matrix elements when interpreting the intensity in dichroic TR-ARPES.

\begin{figure*} [t!]
\begin{center}
\includegraphics[width=1\textwidth]{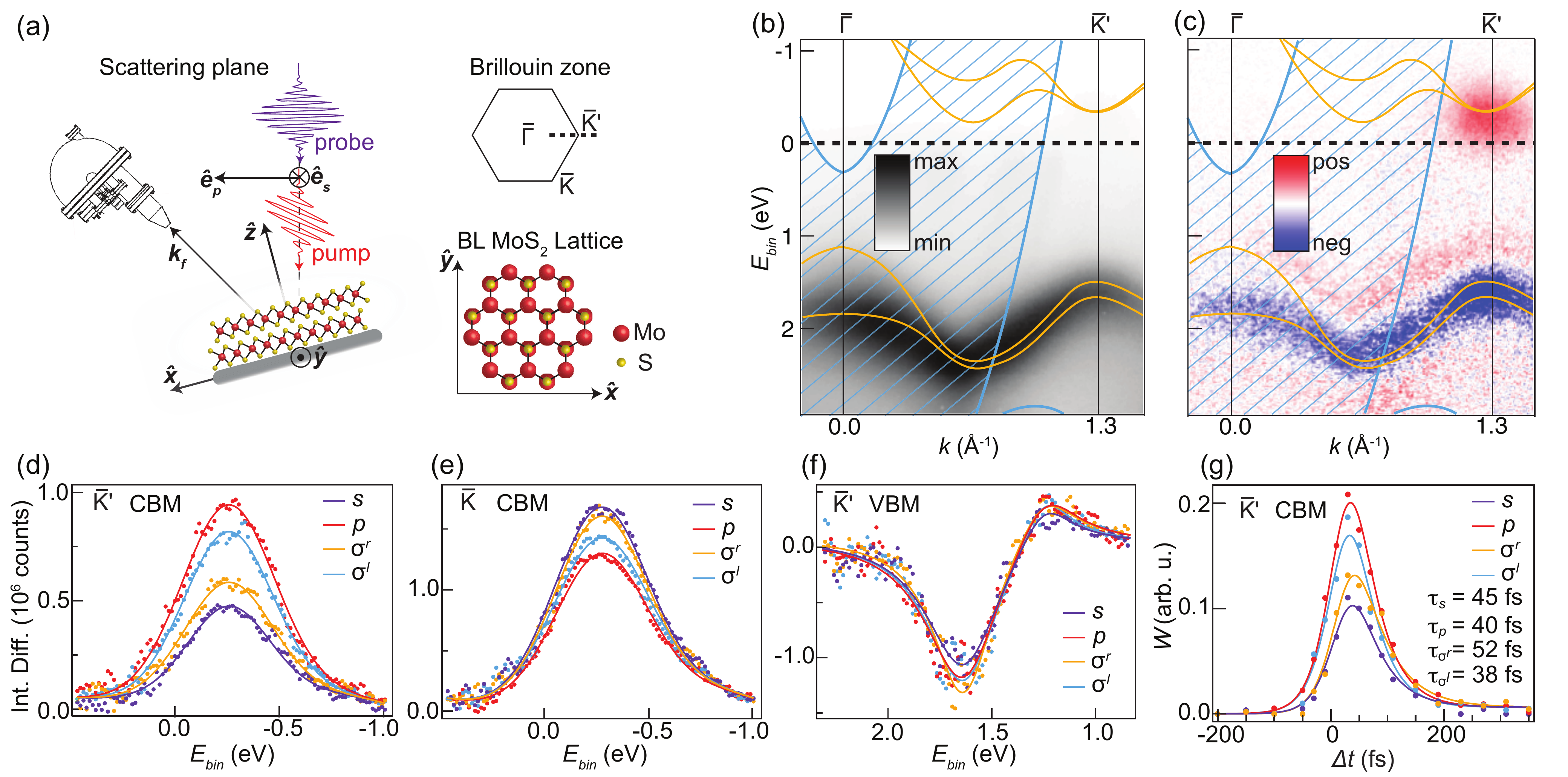}
\caption{(a) TR-ARPES setup: Geometry of the scattering plane with definitions of coordinate system $\{\hat{\vec{x}},\hat{\vec{y}},\hat{\vec{z}}\}$, photoelectron wavevector $\vec{k}_f$ and polarization vectors $\{\hat{\vec{e}}_p,\hat{\vec{e}}_s\}$ of the light pulses. The orientations of the BL MoS$_2$ lattice and the Brillouin zone (BZ) are shown for the measurement geometry. The dashed line on the BZ marks the measurement direction. (b) ARPES intensity along the \gbar-\kbarp~ high symmetry direction before arrival of the pump pulse ($\Delta t<0$). (c) Intensity difference between the spectrum in (b) and one obtained at the peak of the optical excitation at $\Delta t = 40$~fs with a $s$-polarized pump pulse. The dispersion of BL MoS$_2$ has been overlaid (orange curves) together with the bulk continuum of Ag(111) \cite{Dendzik2017} (blue hatched area) in (b)-(c). (d)-(f) EDCs of the intensity difference for different pump polarizations, fitted with Gaussian peaks for the CBM at (d) \kbarp~and (e) \kbar, respectively, as well as (f) for the VBM at \kbarp. (g) Time dependence of the estimated spectral weight $W$ from (d) fitted with a function composed of an exponential rise and a single exponential decay with the given time constants $\tau_i$.}
\label{fig:1}
\end{center}
\end{figure*}

Our BL MoS$_2$ sample is grown on Ag(111) and has predominantly one domain orientation, as determined by X-ray photoelectron diffraction measurements \cite{Baraldi2003,SMAT}. TR-ARPES spectra are collected in the ultra-high vacuum end-station at the Artemis Facility at the Central Laser Facility using the scattering geometry depicted in Fig. \ref{fig:1}(a) \cite{SMAT}. A 32.5~eV probe pulse with a linear polarization fixed parallel to the scattering plane is used, following an optical excitation with a 2~eV pulse. The time resolution is 40~fs and the sample temperature is 300~K.
We tune the polarization of the pump pulse  to obtain arbitrary elliptical polarizations, including linear parallel ($p$) and perpendicular ($s$), with respect to the scattering plane. TR-ARPES measurements have been performed near either  \kbarp~or  \kbar~by an azimuthal rotation of the sample.

The photoemission intensity at a time delay, $\Delta t$, before excitation ($\Delta t<0$) is shown in Fig. \ref{fig:1}(b). The spectrum mainly resembles the bare dispersion of BL MoS$_2$ (orange curves) \cite{He:2014} while the intensity from the Ag(111) bulk states (expected in the blue hatched area) is very faint. The intensity of the global VBM of BL MoS$_2$ (upper band at \gbar)~is strongly suppressed at the energy of the probe pulse, but it is clearly visible at higher photon energies when measured with static ARPES \cite{Hoffmann:2004aa,SMAT}. Figure \ref{fig:1}(c) displays the intensity difference between the equilibrium spectrum in panel~(b) and an excited state spectrum collected at the peak of optical excitation at $\Delta t=40$~fs with a $s$-polarized pump pulse. The dominant features are a strong excitation of electrons around the local CBM at \kbarp~(positive intensity difference) and an excitation of holes that peaks at the local VBM at \kbarp~(negative intensity difference). The red/blue intensity difference associated with the rest of the valence band (VB) is mainly caused by a linewidth broadening of the entire band, which is induced by the pump pulse as observed in TR-ARPES measurements of other materials \cite{Ulstrup:2015bb}. From these data we extract an indirect gap of (1.49$\pm$0.06)~eV from \gbar~to \kbarp~ and a direct gap at \kbarp~of (1.90$\pm$0.04)~eV, which is close to resonant with our 2~eV optical excitation.

\begin{figure} [t!]
\begin{center}
\includegraphics[width=0.49\textwidth]{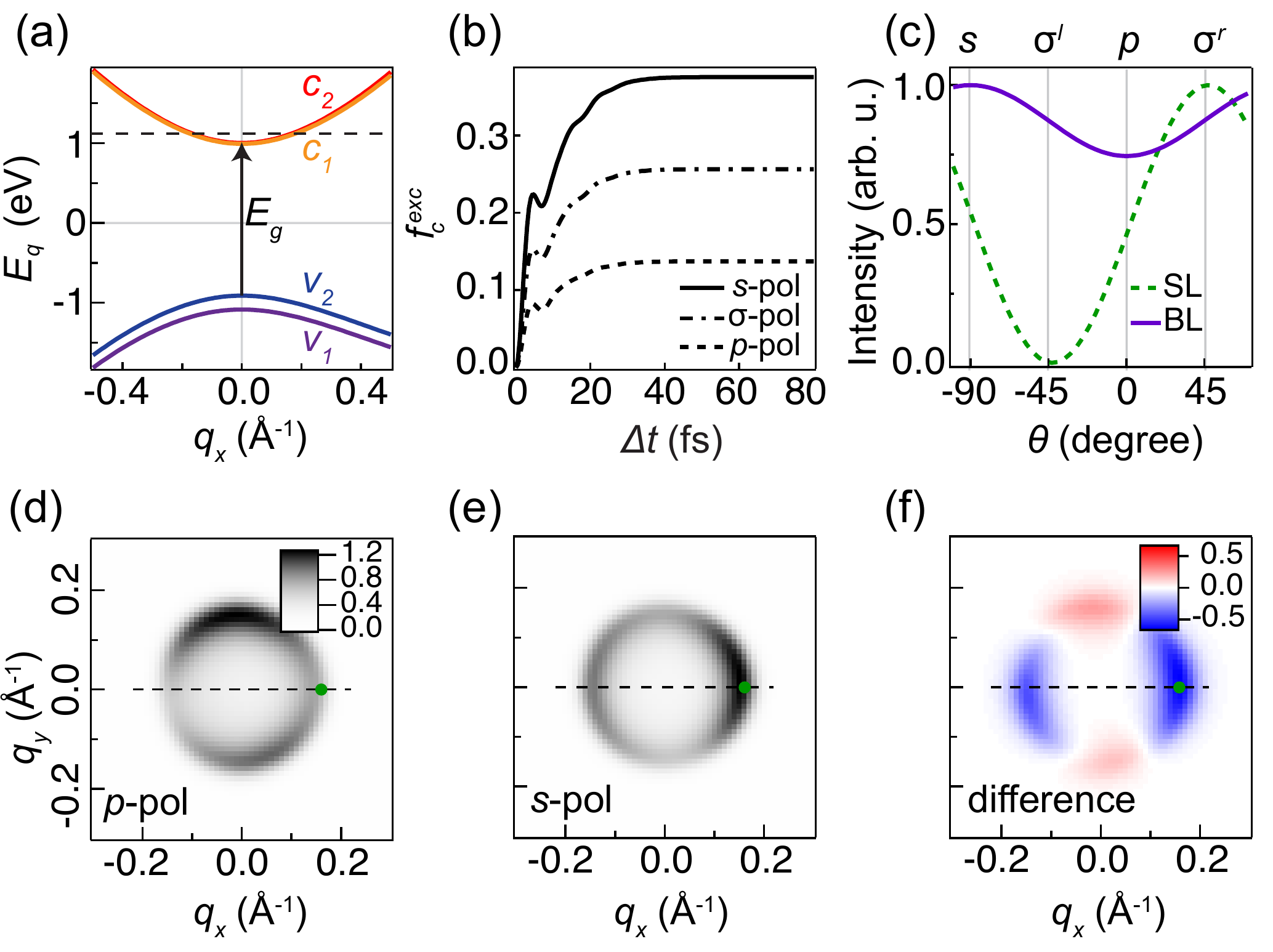}
\caption{(a) BL MoS$_2$ dispersion around \kbar~determined within a $\vec{k} \cdot \vec{p}$ model. The energy split VB states are labeled $v_1$ and $v_2$ and the CB states are labeled $c_1$ and $c_2$. The direct band gap of $E_g = 1.9$ eV is marked by an arrow. (b) Optically induced population in the CB states obtained at $\vec{q} = (0.16,0)$~\AA$^{-1}$ (indicated by the green dot in (d)-(f)) by solving the semiconductor Bloch equations for $p$-, $\sigma$- and $s$-polarized light. (c) Photoemission intensity calculated at $q_y = 0$ and $q_x$-averaged from -0.22 to 0.22~\AA$^{-1}$, as shown by the dashed black line in (d)-(f), as a function of pump pulse polarization $\theta$ for SL MoS$_2$ (dashed green) and BL MoS$_2$ (solid purple). (d)-(e) Momentum-dependent photoemission intensity calculated for (d) $p$- and (e) $s$-polarized pump pulses. The color scale in (d) also applies to (e). (f) Intensity difference obtained by subtracting the data in (e) from that in (d). The constant energy cuts in (d)-(f) were obtained at $E_q=1.15$~eV as shown via a dashed horizontal line in (a).}
\label{fig:2}
\end{center}
\end{figure}

Energy distribution curves (EDCs) of the intensity difference in the CBM averaged over a momentum range from -0.2 to 0.2~\AA$^{-1}$ around \kbarp~are shown with Gaussian fits for $s$- and $p$-polarized pump pulses in Fig. \ref{fig:1}(d). EDCs are also shown for polarizations generated by rotations of the half-waveplate midway between $s$- and $p$-polarizations, which we assume to be circular and therefore label $\sigma^{l}$ and $\sigma^{r}$ \cite{SMAT}.  A strong change is visible between the intensity difference spectra of $s$- and $p$-polarized pulses, while a smaller change is seen between $\sigma^{l}$- and $\sigma^{r}$-polarizations. We quantify dichroism as $\rho_{ij} = (W(i) - W(j))/(W(i) + W(j))$, where $i \neq j$ labels $s$- and $p$- or $\sigma^{l}$- and $\sigma^{r}$-polarization and $W$ represents the corresponding spectral weight, determined as the area of the EDC fits. At \kbarp~we then obtain $\rho_{ps} = 42.4$~\% and $\rho_{\sigma^{l}\sigma^{r}} = 19.7$~\%. A similar EDC analysis for the CBM at \kbar~is presented in Fig. \ref{fig:1}(e) and exhibits a reversal and reduction of the linear dichroism effect with  $\rho_{ps} =-15.2$~\% and likewise $\rho_{\sigma^{l}\sigma^{r}} = -4.7$~\%. EDCs of the VBM at \kbarp~in Fig. \ref{fig:1}(f) do not permit us to clearly distinguish dichroism from the overall noise level in this spectral region. We speculate that ultrafast momentum relaxation of the holes involving the remaining VB states leads to a loss of polarization information of the holes on a faster timescale than we can resolve with our experimental setup.

The time- and polarization-dependence of the spectral weight $W$ at \kbarp~is shown in Fig. \ref{fig:1}(g). In all cases we find that the decay part is well described by single exponentials with the time constants $\tau_i$ given in Fig. \ref{fig:1}(g). The values of $\tau_i$ are similar for all cases and indicate that the carriers are rapidly scattered into the metal substrate \cite{Ulstrup:2015bb,Antonija-Grubisic-Cabo:2015aa}. The excitation signal is detectable up to 300~fs while the dichroic signal exceeds the noise level for only 85~fs.

We seek an explanation of the measured dichroism by calculating the polarization-dependent photoemission intensity from the transiently populated CB around \kbar. A full account of the theory is given in Ref. \citenum{ourprb}. As a simplified expression we use 
\begin{align}
{\cal I}_n(E,\vec{q},\theta) &\propto  |{\cal M}_{n}({\vec{q}})|^2 {\cal A}_n(E,{\vec{q}})f^{exc}_n(\vec{q},\theta),
\label{eq:transphoto}
\end{align}
where ${\cal A}_n$ is the photohole spectral function and $f^{exc}_n$ is the excited state population in the CBs, which we treat as two-fold degenerate and label with the index $n\in\{c_1,c_2\}$ where $c_1$ and $c_2$ are the two states. The wave vector is measured from \kbar~and expressed as $\vec{q} = (q_x,q_y)$ in units of $\sqrt{3}/a$ where $a~=~3.16$~\AA~is the MoS$_2$ lattice constant. The one-electron dipole matrix element (${\cal M}_{n}$) describing photoemission from the CB near \kbar~is determined by the simple assumption that the initial Bloch state ($\psi_{n}$) is matched to a single free-electron final state in the photoemission process. This assumption leads to the expression $\mathcal{M}_n(\vec{q}) \propto \hat{\vec{e}}_p \cdot \vec{k}_f\braket{\vec{k}_f|\psi_{n}(\vec{q})}$, where $\hat{\vec e}_p$ is the polarization unit vector of the probe pulse (see Fig. \ref{fig:1}(a)) and $\vec{k}_f$ is the wavevector of the free electron \cite{Moser:2017,ourprb}. The wave functions and dispersion are determined by diagonalizing a $\vec{k} \cdot \vec{p}$  Hamiltonian given by
\begin{equation}
\hat{\cal H}_{\rm BL}({\vec q},\tau_z,s_z) = \begin{bmatrix} \hat{\cal H}_{\rm SL}(-{\vec q},-\tau_z,s_z) & \hat{\cal H}_{\perp}({\vec q},\tau_z) \\ \hat{\cal H}^\dagger_{\perp}({\vec q},\tau_z) 
& \hat{\cal H}_{\rm SL}({\vec q},\tau_z,s_z) \end{bmatrix},
\label{eq:1}
\end{equation}
where $\hat{\cal H}_{\rm SL}$ is the Hamiltonian for SL MoS$_2$ including trigonal warping effects \cite{Kormanyos:2013,Rostami:2013,Rostami:2016}, and $\hat{\cal H}_\perp$ accounts for the weak interlayer interaction in BL MoS$_2$ \cite{Gong:2013,Kormanyos:2018,SMAT}. Furthermore, $\tau_z=\pm 1$ and $s_z=\pm 1$ denote the valley and spin indices, respectively. The corresponding dispersion relation is presented in Fig. \ref{fig:2}(a).

To simulate the transient CB population generated by the pump pulse we model the electric field as an ultrashort pulse given by $\bm{\mathcal E}(\Delta t)=\hat{\bm\epsilon}(\theta) {\cal E}_0\cos(\omega_0 \Delta t)e^{-(\frac{\Delta t}{\tau_0})^2}$, where $\hat{\vec{\epsilon}}(\theta) = \hat {\vec e}_p \cos{\theta} + \textrm{i}\, \hat {\vec e}_s \sin{\theta}$ is the polarization of the pump pulse with unit vectors $\hat{\vec e}_p$ and $\hat{\vec e}_s$ defined according to the scattering geometry in Fig.~\ref{fig:1}(a). Experimental values are used as input for the electric field strength ${\cal E}_0 = 0.87$~V/nm (determined by the measured pump spot size and fluence), the pulse energy $\hbar \omega_0 = 2.0$~eV and the pulse duration $\tau_0 = 30$~fs. We then solve the semiconductor Bloch equations \cite{haug2004quantum} as described in detail in Ref. \citenum{ourprb}, and obtain the excited CB population shown in Fig. \ref{fig:2}(b). There is no decrease of the excited state population in time as we have neglected any relaxation processes. As we shall see below, the dichroism in the excited state vanishes exactly at \kbar. The plot has therefore been made at the point $\vec{q} = (0.16,0)$~\AA$^{-1}$, which clearly reveals a strong polarization-dependence of the population at finite $\vec{q}$. The noticeable cusps in the early stages of the excitation are strongly dependent on the model parameters and are not considered in further detail here since they can not be resolved in our experiment.  
For the resonance condition $|\hbar\omega_0-E_g|\ll \hbar/\tau_0$ around \kbar~(i.e. small $q$), we can write the population as  \cite{ourprb} 
\begin{align}
f^{exc}_n(\vec{q},\theta) \approx   \left(\frac{\sqrt{\pi}\tau_0 e{\cal E}_0}{4E_g}\right)^2\sum_{m}  |{\cal M}_{nm}(\vec{q},\theta)|^2,
\end{align} 
where $E_g = 1.9$ eV is the measured direct band gap, the sum is over the two valence bands, $m\in\{v_1,v_2\}$, and $\mathcal{M}_{nm}$ is the velocity matrix element describing the interband transition and is given as $\mathcal{M}_{nm}(\vec{q},\theta) = \braket{\psi_{n}(\vec{q}) | \hat{\vec{\epsilon}}(\theta) \cdot {\bm \nabla}_{\bm q} \hat{\cal H}_{\rm BL} | \psi_{m}(\vec{q})}$. Note that, since we only consider the fully excited state and neglect relaxation in the model, we drop the dependence on $\Delta t$. After summing over the possible transitions from the VB to the two-fold degenerate CB, we can formally write the excited state population as \cite{ourprb}
\begin{align}
f^{exc}(\vec{q},\theta)\propto 
  1  + f_{\rm lin}(\vec{q})\cos(2\theta)+ f_{\rm circ}(\vec{q}) \sin(2\theta),
\label{eq:2}
\end{align}
where $f_{\rm lin}$ and $f_{\rm circ}$ are spin-, valley- and momentum-dependent pre-factors that determine the weight between the linear ($\cos(2\theta)$) and circular ($\sin(2\theta)$) dichroism terms.

We calculate the intensity using Eq. (\ref{eq:transphoto}) based on the assumption that the photoemitted electrons from the CB at $\vec{q}$ stem exclusively from pumping into this state and not from the decay of higher lying states. In our case, this is a justified assumption because of the resonant pumping condition. Moreover, we neglect many-body interactions and therefore disregard any recombination effects. Since our TR-ARPES spectra were collected along the \gbar-\kbar~direction and the dichroism was extracted for a $k$-range covering the CBM, we calculate the  intensity resulting from the resonant excitation to the CB at $q_y=0$ and sum over $q_x$ from -0.22 to 0.22~\AA$^{-1}$ for all possible pump pulse polarization angles, as shown for both SL and BL MoS$_2$ in Fig. \ref{fig:2}(c). Our model recovers strong circular dichroism in SL MoS$_2$ (valley-polarization), but with a slightly asymmetric polarization-dependence due to a finite contribution from the linear dichroism term in Eq. (\ref{eq:2}) at finite $\vec{q}$. In BL MoS$_2$ the circular dichroism term is absent, due to inversion symmetry, and we only observe the $\cos(2\theta)$ dependence.

The complete $(q_x,q_y)$-dependent photoemission intensity in the CB of BL MoS$_2$ is shown for $p$- and $s$-polarized excitations in Figs. \ref{fig:2}(d)-(e), respectively. The intensity  difference between these cuts is shown in Fig. \ref{fig:2}(f), where the blue (red) color signifies a negative (positive) sign corresponding to a momentum-dependent linear dichroism effect. Neglecting spin-orbit coupling and trigonal warping, the leading intralayer contribution to this momentum-resolved linear dichroism is 
\begin{align}\label{eq:abc}
f_{\rm lin}(\vec{q})\approx 2 \frac{t^2_1-2E_0 E_g}{E^2_g} q^2 \cos(2\phi),
\end{align}
where $t_1$ is the intralayer effective hopping parameter which we set equal to 2.0 eV, $\phi=\arctan(q_y/q_x)$ is the azimuthal angle of vector $\bm q$ and $E_0=3\hbar^2/4\mu a^2$ with the reduced mass of electron-hole pairs given as $\mu~\sim~0.15m_0$. The interlayer coupling term ($\hat{\cal H}_\perp$) in Eq.~(\ref{eq:1}) cancels after summing over the optical transitions to the two-fold degenerate CB. Furthermore, Eq.~(\ref{eq:abc}) shows that linear dichroism only appears at finite momentum ($q$), that it cancels when azimuthally averaging over momentum and that the modulation depends on the intralayer interaction strength $t_1$. Finally, circular dichroism is absent in a perfectly inversion-symmetric BL when neglecting external influences, because $f_{\rm circ}(\vec{q})\approx \tau_z (-\tau_z)$ for the bottom (top) layer, and thus $\tau_z + (-\tau_z) =0$ when summing over layers.

\begin{figure} [t!]
\begin{center}
\includegraphics[width=0.49\textwidth]{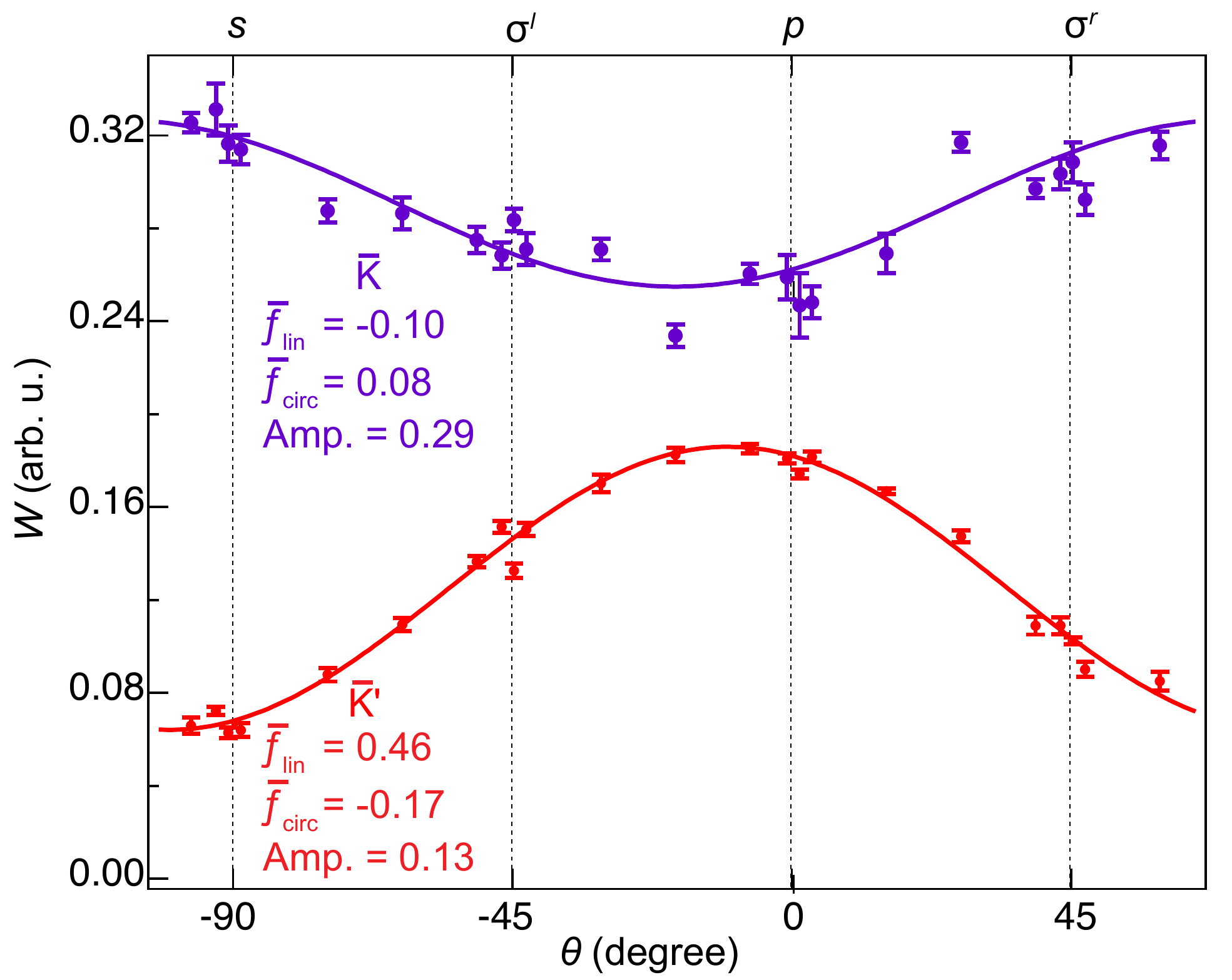}
\caption{Spectral weight $W$ (markers) in the CB as a function of polarization angle $\theta$ measured at \kbarp~(red) and \kbar~(purple). Curves represent fits of the energy- and \textit{k}-integrated photoemission intensity incorporating our model for the excited-state population in Eq. (\ref{eq:2}) with the given pre-factors and amplitude (Amp.).}
\label{fig:3}
\end{center}
\end{figure}

The general form of the occupation function written in Eq.~(\ref{eq:2}) shows that the intensity in polarization-dependent TR-ARPES spectra can be decomposed into linear and circular dichroism contributions. We explore this idea further by fitting the measured $\theta$-dependent spectral weight ($W$) at the peak of the optical excitation to an energy- and $k$-integrated form of the intensity in Eq. (\ref{eq:transphoto}) with the expression in Eq.~(\ref{eq:2}) implemented. The spectral weight and the results of the fitting are presented at \kbarp~and \kbar~in Fig. \ref{fig:3}. The fits provide the amplitudes $\bar f_{\rm lin}$ and $\bar f_{\rm circ}$, which represent an average over the $k$-range used to determine $W$. The values are stated in Fig. \ref{fig:3} and reveal that both the  $\cos(2\theta)$- and $\sin(2\theta)$-terms are significant in our sample. The $\cos(2\theta)$-dependence is in agreement with the model around \kbar~for BL MoS$_2$. The sign-change between \kbarp~and \kbar~of the $\cos(2\theta)$-terms is not explained by our simple model and may arise due to a strong anisotropy of the photoemission matrix element (${\cal M}_{n}$) as described in Ref. \citenum{ourprb}. Many-body interactions such as carrier scattering between the two valleys and recombination can lead to additional contributions to the intensity that are neglected in our non-interacting model. The finite $\bar f_{\rm circ}$ could originate from symmetry breaking due to the substrate or due to the surface-layer sensitivity of ARPES \cite{Riley:2014}. Additionally, the TR-ARPES signal might be influenced by a minor presence of single-domain SL MoS$_2$ areas on the sample. 

Our results show that unexpected dichroism can emerge in energy- and momentum-resolved measurements of transiently populated states even in inversion symmetric materials with an isotropic electronic structure. We have provided a simple model to deconvolve the linear and circular components of the dichroism, which can be linked to the intralayer single-electron hopping in the case of BL MoS$_2$. 
We believe that future TR-ARPES experiments with access to energy-, momentum-, time- and polarization-degrees of freedom in the full BZ will make it possible to uncover the role of single-particle effects and carrier-carrier scattering in shaping the dichroic signal in multiple valleys. Extending these methods to other material systems may lead to the possibility of observing complex topological properties of unoccupied states such as the Berry curvature, in addition to valley dependent selection rules. \\

We thank Phil Rice and Alistair Cox for technical support during the Artemis beamtime. We gratefully acknowledge funding from VILLUM FONDEN through the Young Investigator Program (Grant. No. 15375) and the Centre of Excellence for Dirac Materials (Grant. No. 11744), the Danish Council for Independent Research, Natural Sciences under the Sapere Aude program (Grant No. DFF-4002-00029 and DFF-6108-00409) and the Aarhus University Research Foundation. Access to the Artemis Facility was funded by STFC. H.R. acknowledges the support from the Swedish Research Council (VR 2018-04252). I.M. acknowledges financial support by the International Max Planck Research School for Chemistry and Physics of Quantum Materials (IMPRS-CPQM).

\clearpage
\thispagestyle{empty}

\section{Supporting Information}

\subsection{Growth and photoelectron diffraction of single-domain BL MoS$_2$}

The growth and characterization of MoS${_2}$ samples were carried out at the SuperESCA beamline \cite{Baraldi2003_2} at the Elettra Synchrotron radiation facility in Trieste, Italy.  The Ag(111) substrate was cleaned by repeated cycles of Ar$^+$ ion sputtering at 1 keV and annealing up to 823 K for 10 minutes.  X-ray photoemission spectroscopy (XPS) spectra were used to check the cleanliness of the substrate surface and did not detect any trace of contaminants, with a sensitivity less than 1$\%$. The long range order of the surface was verified with low energy electron diffraction (LEED), which showed a well defined 1x1 pattern.

The growth was monitored and guided by means of fast XPS and consisted of dosing Mo from a home built evaporator, while keeping the Ag substrate at 823 K and dosing H$_2$S (nominal purity 99.8~\%) through a leak valve at background pressure of 2$\times$10$^{-7}$~mbar. The Mo deposition rate was estimated by means of a quartz microbalance and amounted to ~3$\times$10$^{-3}$ ML/minute (where a monolayer (ML) is here defined as the surface atomic density of the Ag(111)). Therefore, the total amount of Mo deposited in 15600~s was $\approx$0.78~ML. 
The final BL MoS$_2$ coverage was determined by looking at the attenuation of the Ag 3$d$ core level and determined to be $\approx$40~\% assuming only BL MoS$_2$.

High-resolution  Mo 3$d$ core level spectra were measured at normal emission on the as-grown MoS$_2$ sample fixed at room temperature, using a photon energy of 360 eV. The overall energy resolution was less than 50 meV. All binding energies are referenced to the Fermi level of the Ag substrate. XPS measurements of the Mo 3$d$ and S 2$p$ core levels were necessary to optimize the growth parameters to obtain pure MoS$_2$. The formation of partially sulfided Mo would be evident by XPS peaks at binding energies lower than the ones attributed to MoS$_2$ and could be actively avoided by tuning the growth parameters during synthesis. \citep{Baker_1999_2}.

\begin{figure*}
\includegraphics[width=0.9\linewidth]{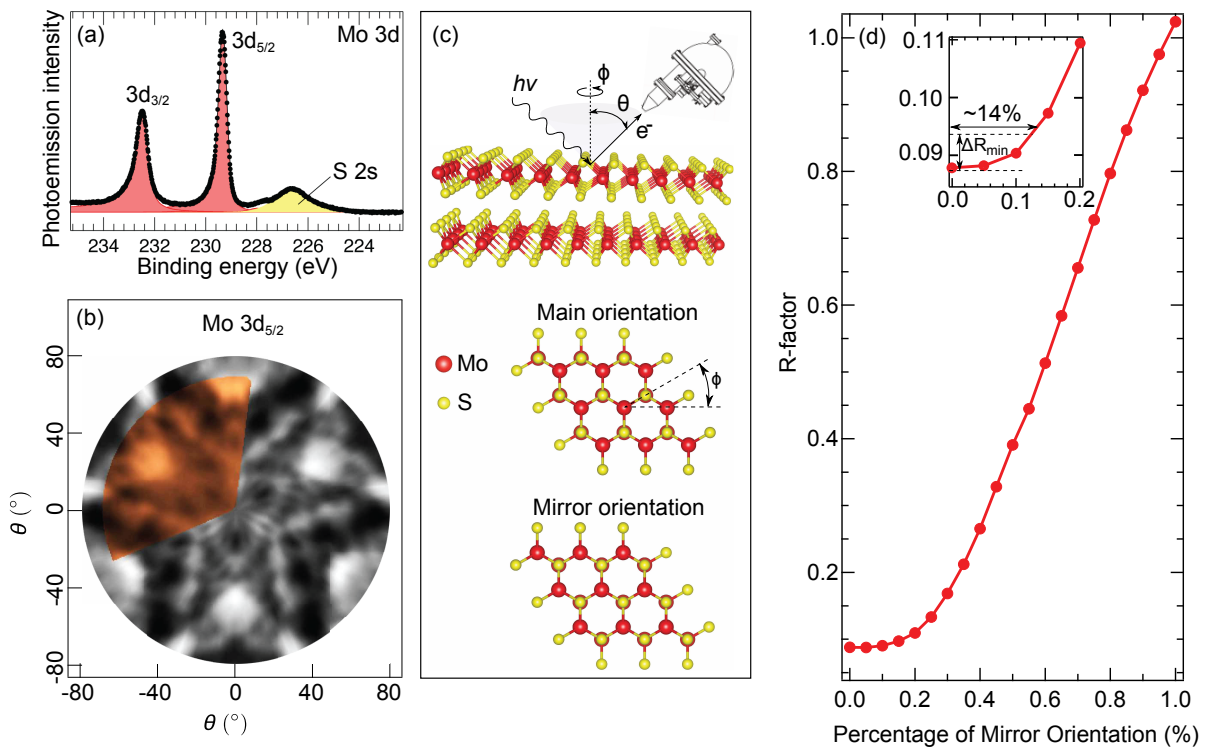}
\caption{\textbf{Characterization of BL MoS$_2$ orientation using photoelectron diffraction.} (a) High resolution Mo 3$d$ core level spectra   together with the spectral contributions resulting from the peak fit analysis (photon energy h$\nu$=360 eV). (b) Azimuthal equidistant polar projection (AEPP) of the integrated photoemission intensity modulation as a function of emission polar ($\theta$) and azimuthal ($\phi$) angles for the Mo 3$d_{5/2}$ core level (photon energy h$\nu$ = 480 eV; E$_k$  $\sim$ 251 eV). The colored sector is the experimental data while the simulated pattern is shown in grey.  (c) Experimental geometry for XPD with  analyzer and  X-ray beam lying in the horizontal plane at an angle of 70$^\circ$ from each other.(d) R-factor behavior as a function of the percentage of mirror orientation. The inset shows a magnification around the minimum of the R-factor, with the dashed lines indicating the confidence interval.}
 \label{Fig:S1}
\end{figure*}

Fig. \ref{Fig:S1}(a)  shows the high resolution Mo 3$d$ core level spectra together with the spectral contributions resulting
from the peak fit analysis.
From this analysis it is not possible to ascertain the bilayer character of MoS$_2$, since the Mo 3$d_{5/2}$ core level only shows a single component. However, peak components are not a decisive indicator of thickness as the Mo atoms could be in chemically similar environments. For this reason the presence of the BL MoS$_2$ was verified through the ARPES measurements, described in Supplementary Section 2. In order to demonstrate the single-orientation character of the top MoS$_2$ layer in the analysis below, we assumed that only BL MoS$_2$ regions are present on the surface (i.e. there are no monolayer or trilayer regions). 

The X-ray photoelectron diffraction (XPD) pattern for Mo 3$d_{5/2}$, shown in Fig. \ref{Fig:S1}(b), was used to determine the orientation of the BL MoS$_2$.
The pattern was obtained by collecting about 40 azimuthal scans
over a wide azimuthal sector of 120$^{\circ}$, from normal ($\theta$=0$^{\circ}$) to grazing ($\theta$=70$^{\circ}$) emission, 
measuring the Mo 3$d_{5/2}$ core-level region. The intensity $I(\theta$, $\phi)$ of each component resulting
from the fit of the spectra --\textit{i.e.}, the area under the photoemission line --was
extracted.
The resulting XPD pattern is the azimuthal equidistant polar projection (AEPP) of the modulation function $\chi$ defined as
\begin{equation}
\chi=\frac{I(\theta, \phi) - I_0(\theta)}{I_0(\theta)},
\label{eq:chi}\end{equation}
 where $I_0$($\theta$) is the average intensity for each azimuthal scan at polar angle $\theta$. The evaluation of the percentage of main and mirror domains was performed by comparing measured XPD patterns to multiple scattering simulations using the program package for Electron Diffraction in Atomic Clusters (EDAC) \cite{Garcia_2001_2}.
The simulated atomic structure is reported in Fig. \ref{Fig:S1}(c) and shows two MoS$_2$ layers arranged according to the so called 2H structure, where Mo and S atoms of the upper layer sits on top S and Mo atoms of the lower layer,  respectively.
In the simulations we only accounted for the BL MoS$_2$, and neglected the Ag substrate, which is appropriate because of the lack of a specific local adsorption configuration of  MoS$_2$ on the substrate due to the lattice mismatch with Ag(111).
For simplicity, the domain sizes are assumed to be sufficiently large so that we can neglect boundary effects \cite{Gronborg:2015aa_2,Dendzik:2015aa_2}.  The  possibility of two coexisting mirror-domain orientations is taken into account as an incoherent superposition of the intensities that would be expected to arise from the two layer orientations depicted in Fig. \ref{Fig:S1}(c).
The lattice parameter and the S-S inter-plane distance used in the simulations are  3.17~\AA~in accordance with the values reported in Ref. \cite{Bana2018_2} for MoS$_2$ on Au(111). 
The agreement between simulations and  experimental results was quantified by computing the reliability factor $(R)$ \cite{xpd_rfac_2}
\begin{equation}R = \frac{\sum_{i} (\chi_{\text{exp}, i} - \chi_{\text{sim}, i})^2}{\sum_{i} ({\chi^2}_{\text{exp}, i} + {\chi^2}_{\text{sim}, i})},\label{eq:R}\end{equation} where  $\chi_{\text{sim}, i}$ and  $\chi_{\text{exp}, i}$ are the simulated and the experimental modulation functions for each emission angle $i$.
Fig. \ref{Fig:S1}(d) shows a minimum of the R-factor when 
only the main orientation is considered in the simulation.

The estimation of the accuracy on the evaluation of the percentage of mirror oriented domains was deduced from the R-factor confidence interval defined as  \cite{Pendry_1980_2}
\begin{equation}
\Delta R_{\text{min}}=R_{\text{min}}\sqrt{\frac{2}{N}},
\label{eq:ConfidenceInterval}
\end{equation}
where $R_{\text{min}}$ is the minimum R-factor value and $N$ is the number of well-resolved peaks in the XPD pattern $(N\sim250)$
\textit{i.e.}, the approximate number of peaks, considering the whole 40
azimuthal scans acquired at different polar emission angles ($\theta$).
From this analysis it turns out that less than 15$\%$ of the BL MoS$_2$ domains assume the mirror orientation.

\subsection{High resolution ARPES measurements}

In addition to XPD and TR-ARPES measurements, static ARPES measurements of the same BL MoS2 sample were acquired at the SGM3 ARPES endstation at the ASTRID2 synchrotron in Aarhus, Denmark \cite{Hoffmann:2004aa_2}. The sample was annealed to 770~K for 10 min to remove adsorbates. Spectra were collected at a variety of photon energies ranging from 31 to 120~eV. The total energy and angular-resolutions were 40~meV and 0.2$^{\circ}$, respectively, and the sample temperature was kept at 70 K throughout the measurements.

\begin{figure} [t!]
\begin{center}
\includegraphics[width=0.49\textwidth]{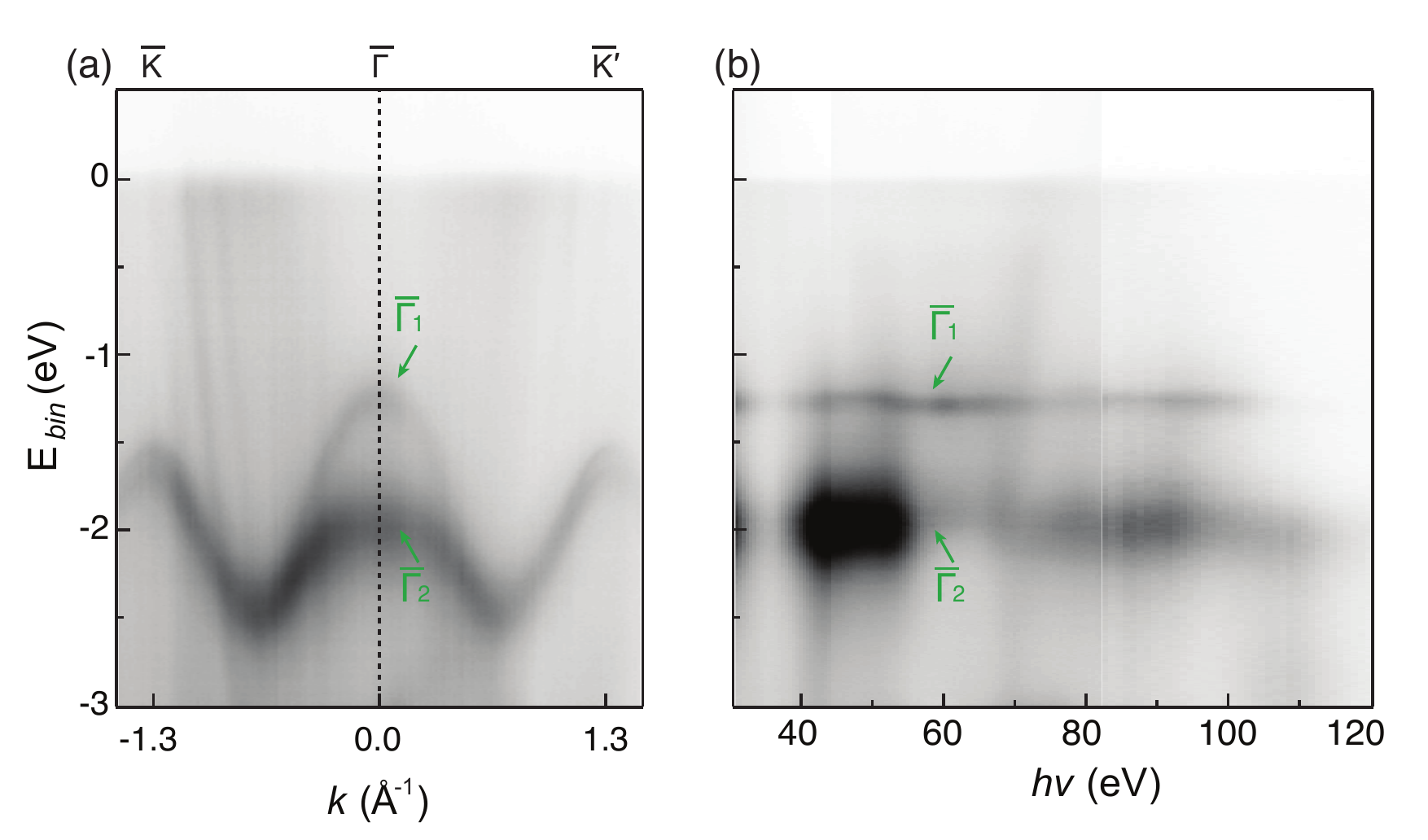}
\caption{\textbf{Static ARPES measurements of BL MoS$_2$.} (a) ARPES intensity along the high symmetry \kgk~direction of BL MoS$_2$ on Ag(111) acquired at a photon energy of 80 eV. (b) Photon energy dependence of \gbar$_1$ and \gbar$_2$ bands at the \gbar-cut marked by a dashed line in (a).}
\label{Fig:S2}
\end{center}
\end{figure}

The ARPES spectrum shown in Fig. \ref{Fig:S2}(a), obtained at a photon energy of 80~eV, confirms the formation of a bilayer by the appearance of the two characteristic bands at \gbar~originating from out-of-plane coupling of S $p_z$ orbitals, exhibiting a splitting of (720 $\pm$ 10)~meV consistent with theoretical calculations \cite{He:2014_2}. Additionally, we observe the expected spin-splitting of (130 $\pm$ 5)~meV at \kbar. The bands at \gbar, labeled \gbar$_1$ and \gbar$_2$, do not disperse with photon energy, as shown in Fig. \ref{Fig:S2}(b), demonstrating that both states have 2D character. The observed splittings at \gbar~and \kbar~as well as the bandwidth of the topmost valence band measured along the \kbar-\gbar~direction are consistent with calculated bands for the 2H-stacking of BL MoS$_2$ \cite{He:2014_2}.

\subsection{Experimental details of TR-ARPES experiment}

The BL MoS$_2$ on Ag(111) crystal was transported from ASTRID2 through air and inserted into the ultra-high vacuum end-station at Artemis, and annealed to 620~K to remove any adsorbed surface contaminants. The sample was then kept at room temperature throughout the measurements.

A 1 kHz Ti:sapphire amplified laser with a pulse duration of 30 fs, a wavelength of 785 nm and an energy per pulse of 12~mJ was used to generate the pump and probe pulses. Probe pulses with a photon energy of 32.5~eV were achieved using high harmonic generation, by the focusing of a part of the fundamental laser energy inside a jet of Ar gas. The energy of the pump pulses was tuned to 2.00~eV (621~nm) using an optical parametric amplifier (HE-TOPAS) followed by a frequency mixing stage. The fluence of the pump pulse was kept around 3~mJ/cm$^{2}$ in order to optimize the excitation density while avoiding space-charge effects in the photoemission signal. The time delay between the two pulses was varied using a mechanical delay line. The energy, angular, and time resolution were 400~meV, 0.3$^{\circ}$, and 40~fs, respectively.

\begin{figure} [t!]
\begin{center}
\includegraphics[width=0.49\textwidth]{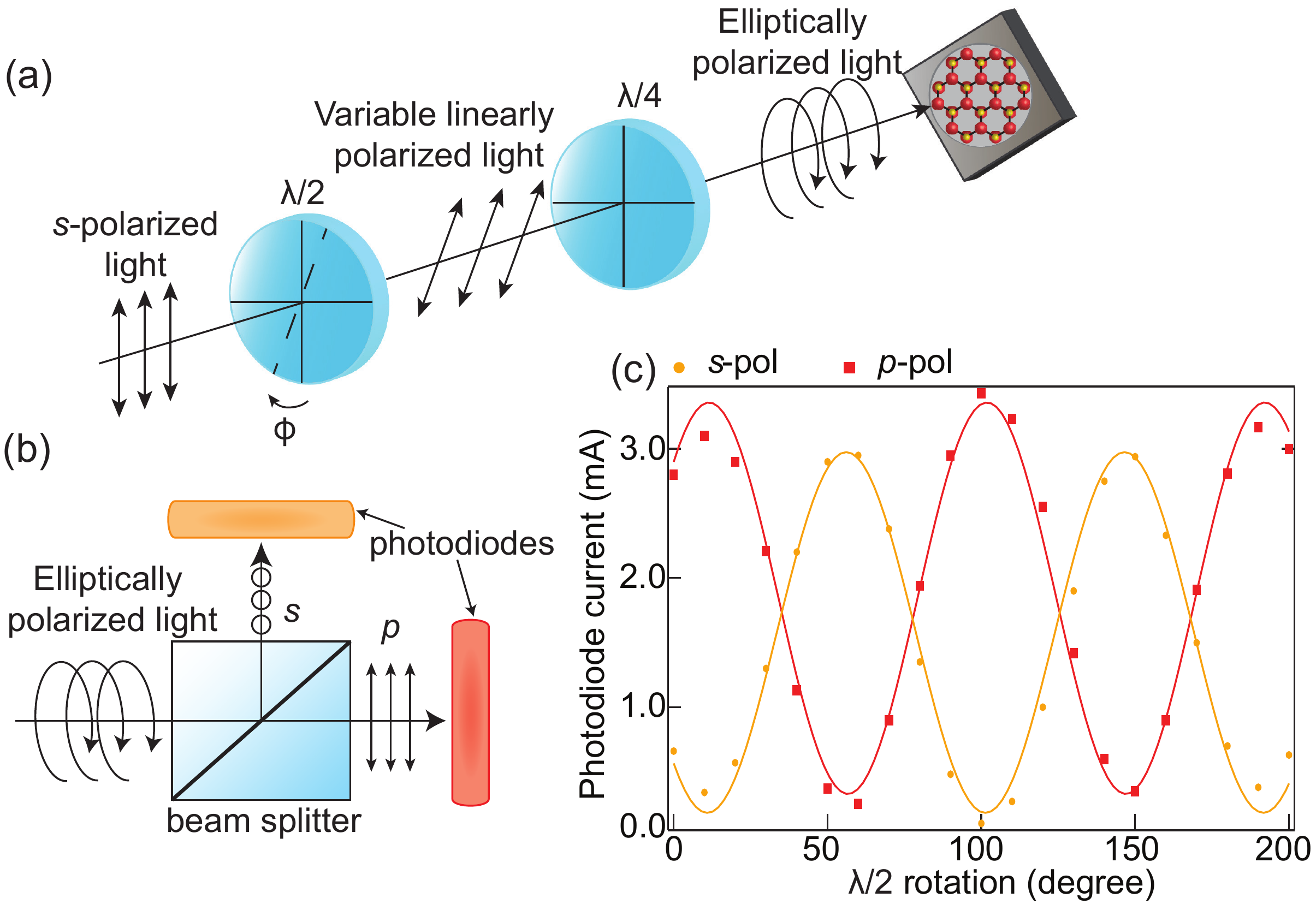}
\caption{\textbf{Calibration of pump pulse polarization.} (a) A rotatable half-waveplate ($\lambda$/2) and quarter-waveplate ($\lambda$/4) inserted into the path of the pump pulse to convert the incoming $s$-polarised light to any elliptical polarization. (b) Polarizing beam splitter introduced after the sample position to measure the $s$- and $p$-polarization components as a function of the half-waveplate angle. (c) The resulting photocurrent (markers) with fits (curves) as a function of $\phi$ allowing us to determine the relative angles for $s$-, $\sigma^{l}$-, $p$- and $\sigma^{r}$-polarizations.}
\label{Fig:S3}
\end{center}
\end{figure}

The elliptical polarization of the pump pulses was tuned using a motorized half-waveplate followed by a fixed quarter-waveplate (Fig. \ref{Fig:S3}(a)). The beam was then deflected on a silver-coated mirror before hitting the sample. The polarization angles were calibrated by moving the sample out of the beam path and sending the beam through a window flange on the chamber and into a polarizing beam splitter, as shown in Fig. \ref{Fig:S3}(b). The resulting photodiode current was fit using a sinusoidal function to extract the peak positions for $s$- and $p$-polarized light ($56^{\circ}$ and $102 ^{\circ}$ respectively). The conversion between waveplate angle and polarization angle, as defined in the main paper, is then $\theta= 2 (\phi - 56^{\circ})$, with the factor of two originating from the doubling of the angle in the half-waveplate. 
We do not explicitly measure the phase of the incident pump beam at the points midway between $s$- and $p$-polarized light, but the quarter-waveplate in the setup would be expected to generate circular polarization at these points, and thus we label them $\sigma^{l}$ and $\sigma^{r}$.  Whether the polarization at these points is strictly circular does not affect our conclusions about intensity differences at $s$- and $p$-polarizations, which are the focus of this study.

\subsection{Equilibrium and intensity difference spectrum}

\begin{figure} [h!]
\begin{center}
\includegraphics[width=0.49\textwidth]{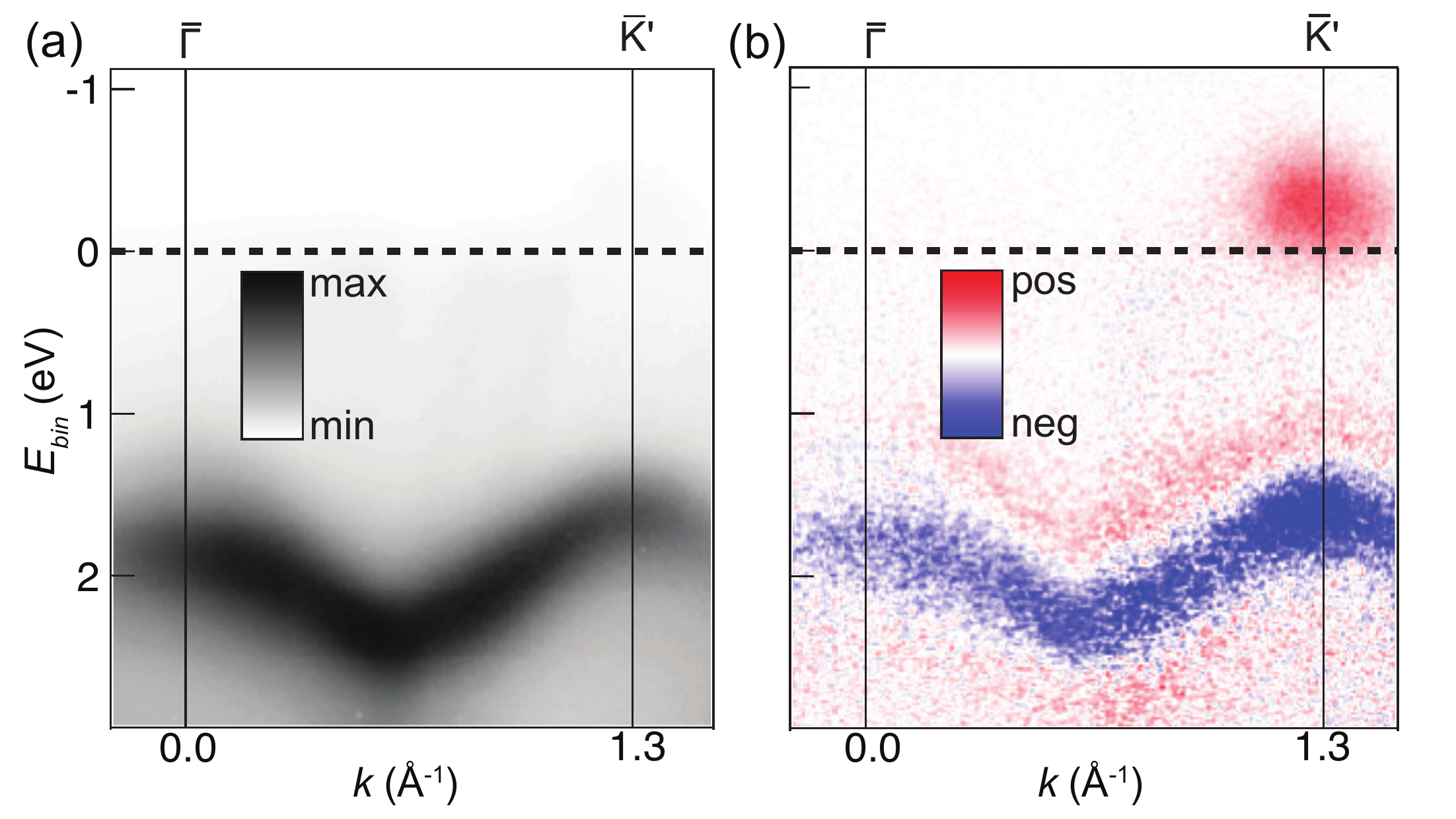}
\caption{\textbf{TR-ARPES measurements.} (a) ARPES intensity along the \kgk~high symmetry direction before arrival of the pump pulse ($t < 0$). (b) Intensity difference between the spectrum in (a) and one obtained at the peak of the optical excitation at $t = 40$ fs with a $s$-polarized pump pulse.}
\label{Fig:S4}
\end{center}
\end{figure}

Duplicates of Figs. 1(b)-(c) in the main text are given in Figs. \ref{Fig:S4}(a)-(b) without the overlaid bilayer MoS$_2$ band structure and Ag(111) bulk continuum lines. 

\subsection{Setup of BL MoS$_2$ Hamiltonian and $\bm k\cdot\bm p$ parameters}
The low-energy Hamiltonian of SL MoS$_2$ reads as $\hat{\cal H}_{\rm SL}= {\vec d} \cdot \hat{\bm\sigma}+[
\lambda_I\tau_z s_z+\alpha q^2]\hat I + \hat {\cal H}_{\rm TW}$, where $\hat{\bm\sigma} = (\hat\sigma_x,\hat\sigma_y,\hat\sigma_z)$ are the Pauli matrices in the band basis, ${\vec d} =  (t_1 \tau_z q_x , t_1 q_y, \Delta+\lambda\tau_z s_z+\beta  q^2)$ and $\hat{{\cal H}}_{\rm TW}$ represents the trigonal warping term that is given in Ref. \cite{ourprb_2}. The interlayer part of the Hamiltonian is given by $\hat{\cal H}_\perp={\rm diag}[t'_\perp (\tau_z q_x+iq_y),t_\perp]$, with $t_\perp=0.045$~eV and $t'_\perp=0.0387$~eV~\cite{Gong:2013_2,Kormanyos:2018_2}. We write $\vec{q} = (q_x,q_y)$ in units of $1/a_0=\sqrt{3}/a$, where $a~=~3.16$~\AA~is the MoS$_2$ lattice constant. Considering electron and hole effective masses, we obtain $\alpha= \hbar^2/4\mu' a^2_0$ and $\beta = E_0 - t^2_1/E_g$ 
with $E_0=\hbar^2/4\mu a^2_0$, where $\mu=m_e m_h/(m_h+m_e)\sim0.15m_0$ and $\mu'=m_e m_h/(m_h-m_e)\sim1.01m_0$. Notice that $2\Delta= E_g-\lambda_c+\sqrt{t^2_\perp+\lambda^2_v}$, in which $E_g=1.9$~eV is the energy gap and $\lambda_c=\lambda_I+\lambda\sim-5.5$~meV and $\lambda_v=\lambda_I-\lambda\sim74.5$~meV are the spin-orbit coupling in the conduction and valence bands, respectively. The intralayer interaction given by the parameter $t_1$ is set equal to 2.0 eV.


\begin{thebibliography}{35}%
\makeatletter
\providecommand \@ifxundefined [1]{%
 \@ifx{#1\undefined}
}%
\providecommand \@ifnum [1]{%
 \ifnum #1\expandafter \@firstoftwo
 \else \expandafter \@secondoftwo
 \fi
}%
\providecommand \@ifx [1]{%
 \ifx #1\expandafter \@firstoftwo
 \else \expandafter \@secondoftwo
 \fi
}%
\providecommand \natexlab [1]{#1}%
\providecommand \enquote  [1]{``#1''}%
\providecommand \bibnamefont  [1]{#1}%
\providecommand \bibfnamefont [1]{#1}%
\providecommand \citenamefont [1]{#1}%
\providecommand \href@noop [0]{\@secondoftwo}%
\providecommand \href [0]{\begingroup \@sanitize@url \@href}%
\providecommand \@href[1]{\@@startlink{#1}\@@href}%
\providecommand \@@href[1]{\endgroup#1\@@endlink}%
\providecommand \@sanitize@url [0]{\catcode `\\12\catcode `\$12\catcode
  `\&12\catcode `\#12\catcode `\^12\catcode `\_12\catcode `\%12\relax}%
\providecommand \@@startlink[1]{}%
\providecommand \@@endlink[0]{}%
\providecommand \url  [0]{\begingroup\@sanitize@url \@url }%
\providecommand \@url [1]{\endgroup\@href {#1}{\urlprefix }}%
\providecommand \urlprefix  [0]{URL }%
\providecommand \Eprint [0]{\href }%
\providecommand \doibase [0]{http://dx.doi.org/}%
\providecommand \selectlanguage [0]{\@gobble}%
\providecommand \bibinfo  [0]{\@secondoftwo}%
\providecommand \bibfield  [0]{\@secondoftwo}%
\providecommand \translation [1]{[#1]}%
\providecommand \BibitemOpen [0]{}%
\providecommand \bibitemStop [0]{}%
\providecommand \bibitemNoStop [0]{.\EOS\space}%
\providecommand \EOS [0]{\spacefactor3000\relax}%
\providecommand \BibitemShut  [1]{\csname bibitem#1\endcsname}%
\let\auto@bib@innerbib\@empty
\bibitem [{\citenamefont {Henderson}\ and\ \citenamefont
  {Imbusch}(2006)}]{Henderson:2006aa}%
  \BibitemOpen
  \bibfield  {author} {\bibinfo {author} {\bibfnamefont {B.}~\bibnamefont
  {Henderson}}\ and\ \bibinfo {author} {\bibfnamefont {G.~F.}\ \bibnamefont
  {Imbusch}},\ }\href@noop {} {\emph {\bibinfo {title} {Optical Spectroscopy of
  Inorganic Solids}}}\ (\bibinfo  {publisher} {Oxford University Press},\
  \bibinfo {year} {2006})\BibitemShut {NoStop}%
\bibitem [{\citenamefont {Eberhardt}\ and\ \citenamefont
  {Plummer}(1980)}]{Eberhardt:1980aa}%
  \BibitemOpen
  \bibfield  {author} {\bibinfo {author} {\bibfnamefont {W.}~\bibnamefont
  {Eberhardt}}\ and\ \bibinfo {author} {\bibfnamefont {E.~W.}\ \bibnamefont
  {Plummer}},\ }\href@noop {} {\bibfield  {journal} {\bibinfo  {journal}
  {Physical Review B}\ }\textbf {\bibinfo {volume} {21}},\ \bibinfo {pages}
  {3245} (\bibinfo {year} {1980})}\BibitemShut {NoStop}%
\bibitem [{\citenamefont {Gierz}\ \emph {et~al.}(2012)\citenamefont {Gierz},
  \citenamefont {Lindroos}, \citenamefont {H{\"o}chst}, \citenamefont {Ast},\
  and\ \citenamefont {Kern}}]{Gierz:2012}%
  \BibitemOpen
  \bibfield  {author} {\bibinfo {author} {\bibfnamefont {I.}~\bibnamefont
  {Gierz}}, \bibinfo {author} {\bibfnamefont {M.}~\bibnamefont {Lindroos}},
  \bibinfo {author} {\bibfnamefont {H.}~\bibnamefont {H{\"o}chst}}, \bibinfo
  {author} {\bibfnamefont {C.~R.}\ \bibnamefont {Ast}}, \ and\ \bibinfo
  {author} {\bibfnamefont {K.}~\bibnamefont {Kern}},\ }\href {\doibase
  10.1021/nl300512q} {\bibfield  {journal} {\bibinfo  {journal} {Nano Letters}\
  }\textbf {\bibinfo {volume} {12}},\ \bibinfo {pages} {3900} (\bibinfo {year}
  {2012})}\BibitemShut {NoStop}%
\bibitem [{\citenamefont {Zhu}\ \emph {et~al.}(2013)\citenamefont {Zhu},
  \citenamefont {Veenstra}, \citenamefont {Levy}, \citenamefont {Ubaldini},
  \citenamefont {Syers}, \citenamefont {Butch}, \citenamefont {Paglione},
  \citenamefont {Haverkort}, \citenamefont {Elfimov},\ and\ \citenamefont
  {Damascelli}}]{Zhu:2013aa}%
  \BibitemOpen
  \bibfield  {author} {\bibinfo {author} {\bibfnamefont {Z.-H.}\ \bibnamefont
  {Zhu}}, \bibinfo {author} {\bibfnamefont {C.~N.}\ \bibnamefont {Veenstra}},
  \bibinfo {author} {\bibfnamefont {G.}~\bibnamefont {Levy}}, \bibinfo {author}
  {\bibfnamefont {A.}~\bibnamefont {Ubaldini}}, \bibinfo {author}
  {\bibfnamefont {P.}~\bibnamefont {Syers}}, \bibinfo {author} {\bibfnamefont
  {N.~P.}\ \bibnamefont {Butch}}, \bibinfo {author} {\bibfnamefont
  {J.}~\bibnamefont {Paglione}}, \bibinfo {author} {\bibfnamefont {M.~W.}\
  \bibnamefont {Haverkort}}, \bibinfo {author} {\bibfnamefont {I.~S.}\
  \bibnamefont {Elfimov}}, \ and\ \bibinfo {author} {\bibfnamefont
  {A.}~\bibnamefont {Damascelli}},\ }\href {\doibase
  10.1103/PhysRevLett.110.216401} {\bibfield  {journal} {\bibinfo  {journal}
  {Phys. Rev. Lett.}\ }\textbf {\bibinfo {volume} {110}},\ \bibinfo {pages}
  {216401} (\bibinfo {year} {2013})}\BibitemShut {NoStop}%
\bibitem [{\citenamefont {Cao}\ \emph {et~al.}(2013)\citenamefont {Cao},
  \citenamefont {Waugh}, \citenamefont {Zhang}, \citenamefont {Luo},
  \citenamefont {Wang}, \citenamefont {Reber}, \citenamefont {Mo},
  \citenamefont {Xu}, \citenamefont {Yang}, \citenamefont {Schneeloch},
  \citenamefont {Gu}, \citenamefont {Brahlek}, \citenamefont {Bansal},
  \citenamefont {Oh}, \citenamefont {Zunger},\ and\ \citenamefont
  {Dessau}}]{Cao:2013}%
  \BibitemOpen
  \bibfield  {author} {\bibinfo {author} {\bibfnamefont {Y.}~\bibnamefont
  {Cao}}, \bibinfo {author} {\bibfnamefont {J.~A.}\ \bibnamefont {Waugh}},
  \bibinfo {author} {\bibfnamefont {X.-W.}\ \bibnamefont {Zhang}}, \bibinfo
  {author} {\bibfnamefont {J.-W.}\ \bibnamefont {Luo}}, \bibinfo {author}
  {\bibfnamefont {Q.}~\bibnamefont {Wang}}, \bibinfo {author} {\bibfnamefont
  {T.~J.}\ \bibnamefont {Reber}}, \bibinfo {author} {\bibfnamefont {S.~K.}\
  \bibnamefont {Mo}}, \bibinfo {author} {\bibfnamefont {Z.}~\bibnamefont {Xu}},
  \bibinfo {author} {\bibfnamefont {A.}~\bibnamefont {Yang}}, \bibinfo {author}
  {\bibfnamefont {J.}~\bibnamefont {Schneeloch}}, \bibinfo {author}
  {\bibfnamefont {G.~D.}\ \bibnamefont {Gu}}, \bibinfo {author} {\bibfnamefont
  {M.}~\bibnamefont {Brahlek}}, \bibinfo {author} {\bibfnamefont
  {N.}~\bibnamefont {Bansal}}, \bibinfo {author} {\bibfnamefont
  {S.}~\bibnamefont {Oh}}, \bibinfo {author} {\bibfnamefont {A.}~\bibnamefont
  {Zunger}}, \ and\ \bibinfo {author} {\bibfnamefont {D.~S.}\ \bibnamefont
  {Dessau}},\ }\href {https://doi.org/10.1038/nphys2685} {\bibfield  {journal}
  {\bibinfo  {journal} {Nature Physics}\ }\textbf {\bibinfo {volume} {9}},\
  \bibinfo {pages} {499 EP } (\bibinfo {year} {2013})}\BibitemShut {NoStop}%
\bibitem [{\citenamefont {Sch\"uler}\ \emph {et~al.}(2019)\citenamefont
  {Sch\"uler}, \citenamefont {Giovannini}, \citenamefont {H\"ubener},
  \citenamefont {Rubio}, \citenamefont {Sentef},\ and\ \citenamefont
  {Werner}}]{Schuler:2019}%
  \BibitemOpen
  \bibfield  {author} {\bibinfo {author} {\bibfnamefont {M.}~\bibnamefont
  {Sch\"uler}}, \bibinfo {author} {\bibfnamefont {U.~D.}\ \bibnamefont
  {Giovannini}}, \bibinfo {author} {\bibfnamefont {H.}~\bibnamefont
  {H\"ubener}}, \bibinfo {author} {\bibfnamefont {A.}~\bibnamefont {Rubio}},
  \bibinfo {author} {\bibfnamefont {M.~A.}\ \bibnamefont {Sentef}}, \ and\
  \bibinfo {author} {\bibfnamefont {P.}~\bibnamefont {Werner}},\ }\href@noop {}
  {\enquote {\bibinfo {title} {Local berry curvature signatures in dichroic
  angle-resolved photoelectron spectroscopy},}\ } (\bibinfo {year} {2019}),\
  \Eprint {http://arxiv.org/abs/arXiv:1905.09404} {arXiv:1905.09404}
  \BibitemShut {NoStop}%
\bibitem [{\citenamefont {Cho}\ \emph {et~al.}(2018)\citenamefont {Cho},
  \citenamefont {Park}, \citenamefont {Hong}, \citenamefont {Jung},
  \citenamefont {Kim}, \citenamefont {Han}, \citenamefont {Kyung},
  \citenamefont {Kim}, \citenamefont {Mo}, \citenamefont {Denlinger},
  \citenamefont {Shim}, \citenamefont {Han}, \citenamefont {Kim},\ and\
  \citenamefont {Park}}]{Cho:2018}%
  \BibitemOpen
  \bibfield  {author} {\bibinfo {author} {\bibfnamefont {S.}~\bibnamefont
  {Cho}}, \bibinfo {author} {\bibfnamefont {J.-H.}\ \bibnamefont {Park}},
  \bibinfo {author} {\bibfnamefont {J.}~\bibnamefont {Hong}}, \bibinfo {author}
  {\bibfnamefont {J.}~\bibnamefont {Jung}}, \bibinfo {author} {\bibfnamefont
  {B.~S.}\ \bibnamefont {Kim}}, \bibinfo {author} {\bibfnamefont
  {G.}~\bibnamefont {Han}}, \bibinfo {author} {\bibfnamefont {W.}~\bibnamefont
  {Kyung}}, \bibinfo {author} {\bibfnamefont {Y.}~\bibnamefont {Kim}}, \bibinfo
  {author} {\bibfnamefont {S.-K.}\ \bibnamefont {Mo}}, \bibinfo {author}
  {\bibfnamefont {J.~D.}\ \bibnamefont {Denlinger}}, \bibinfo {author}
  {\bibfnamefont {J.~H.}\ \bibnamefont {Shim}}, \bibinfo {author}
  {\bibfnamefont {J.~H.}\ \bibnamefont {Han}}, \bibinfo {author} {\bibfnamefont
  {C.}~\bibnamefont {Kim}}, \ and\ \bibinfo {author} {\bibfnamefont {S.~R.}\
  \bibnamefont {Park}},\ }\href {\doibase 10.1103/PhysRevLett.121.186401}
  {\bibfield  {journal} {\bibinfo  {journal} {Phys. Rev. Lett.}\ }\textbf
  {\bibinfo {volume} {121}},\ \bibinfo {pages} {186401} (\bibinfo {year}
  {2018})}\BibitemShut {NoStop}%
\bibitem [{\citenamefont {Xiao}\ \emph {et~al.}(2012)\citenamefont {Xiao},
  \citenamefont {Liu}, \citenamefont {Feng}, \citenamefont {Xu},\ and\
  \citenamefont {Yao}}]{Xiao:2012ab}%
  \BibitemOpen
  \bibfield  {author} {\bibinfo {author} {\bibfnamefont {D.}~\bibnamefont
  {Xiao}}, \bibinfo {author} {\bibfnamefont {G.-B.}\ \bibnamefont {Liu}},
  \bibinfo {author} {\bibfnamefont {W.}~\bibnamefont {Feng}}, \bibinfo {author}
  {\bibfnamefont {X.}~\bibnamefont {Xu}}, \ and\ \bibinfo {author}
  {\bibfnamefont {W.}~\bibnamefont {Yao}},\ }\href@noop {} {\bibfield
  {journal} {\bibinfo  {journal} {Phys. Rev. Lett.}\ }\textbf {\bibinfo
  {volume} {108}},\ \bibinfo {pages} {196802} (\bibinfo {year}
  {2012})}\BibitemShut {NoStop}%
\bibitem [{\citenamefont {Mak}\ \emph {et~al.}(2014)\citenamefont {Mak},
  \citenamefont {McGill}, \citenamefont {Park},\ and\ \citenamefont
  {McEuen}}]{Mak:2014aa}%
  \BibitemOpen
  \bibfield  {author} {\bibinfo {author} {\bibfnamefont {K.~F.}\ \bibnamefont
  {Mak}}, \bibinfo {author} {\bibfnamefont {K.~L.}\ \bibnamefont {McGill}},
  \bibinfo {author} {\bibfnamefont {J.}~\bibnamefont {Park}}, \ and\ \bibinfo
  {author} {\bibfnamefont {P.~L.}\ \bibnamefont {McEuen}},\ }\href@noop {}
  {\bibfield  {journal} {\bibinfo  {journal} {Science}\ }\textbf {\bibinfo
  {volume} {344}},\ \bibinfo {pages} {1489} (\bibinfo {year}
  {2014})}\BibitemShut {NoStop}%
\bibitem [{\citenamefont {Mak}\ \emph {et~al.}(2012)\citenamefont {Mak},
  \citenamefont {He}, \citenamefont {Shan},\ and\ \citenamefont
  {Heinz}}]{makcontrol2012}%
  \BibitemOpen
  \bibfield  {author} {\bibinfo {author} {\bibfnamefont {K.}~\bibnamefont
  {Mak}}, \bibinfo {author} {\bibfnamefont {K.}~\bibnamefont {He}}, \bibinfo
  {author} {\bibfnamefont {J.}~\bibnamefont {Shan}}, \ and\ \bibinfo {author}
  {\bibfnamefont {T.}~\bibnamefont {Heinz}},\ }\href@noop {} {\bibfield
  {journal} {\bibinfo  {journal} {Nature Nanotechnology}\ }\textbf {\bibinfo
  {volume} {7}},\ \bibinfo {pages} {494} (\bibinfo {year} {2012})}\BibitemShut
  {NoStop}%
\bibitem [{\citenamefont {Zeng}\ \emph {et~al.}(2012)\citenamefont {Zeng},
  \citenamefont {Dai}, \citenamefont {Yao}, \citenamefont {Xiao},\ and\
  \citenamefont {Cui}}]{zengvalley2012}%
  \BibitemOpen
  \bibfield  {author} {\bibinfo {author} {\bibfnamefont {H.}~\bibnamefont
  {Zeng}}, \bibinfo {author} {\bibfnamefont {J.}~\bibnamefont {Dai}}, \bibinfo
  {author} {\bibfnamefont {W.}~\bibnamefont {Yao}}, \bibinfo {author}
  {\bibfnamefont {D.}~\bibnamefont {Xiao}}, \ and\ \bibinfo {author}
  {\bibfnamefont {X.}~\bibnamefont {Cui}},\ }\href {\doibase
  10.1038/nnano.2012.95} {\bibfield  {journal} {\bibinfo  {journal} {Nature
  Nanotech.}\ }\textbf {\bibinfo {volume} {7}},\ \bibinfo {pages} {490}
  (\bibinfo {year} {2012})}\BibitemShut {NoStop}%
\bibitem [{\citenamefont {Perfetti}\ \emph {et~al.}(2007)\citenamefont
  {Perfetti}, \citenamefont {Loukakos}, \citenamefont {Lisowski}, \citenamefont
  {Bovensiepen}, \citenamefont {Eisaki},\ and\ \citenamefont
  {Wolf}}]{Perfetti:2007}%
  \BibitemOpen
  \bibfield  {author} {\bibinfo {author} {\bibfnamefont {L.}~\bibnamefont
  {Perfetti}}, \bibinfo {author} {\bibfnamefont {P.~A.}\ \bibnamefont
  {Loukakos}}, \bibinfo {author} {\bibfnamefont {M.}~\bibnamefont {Lisowski}},
  \bibinfo {author} {\bibfnamefont {U.}~\bibnamefont {Bovensiepen}}, \bibinfo
  {author} {\bibfnamefont {H.}~\bibnamefont {Eisaki}}, \ and\ \bibinfo {author}
  {\bibfnamefont {M.}~\bibnamefont {Wolf}},\ }\href {\doibase
  10.1103/PhysRevLett.99.197001} {\bibfield  {journal} {\bibinfo  {journal}
  {Physical Review Letters}\ }\textbf {\bibinfo {volume} {99}},\ \bibinfo
  {pages} {197001} (\bibinfo {year} {2007})}\BibitemShut {NoStop}%
\bibitem [{\citenamefont {Malic}\ \emph {et~al.}(2011)\citenamefont {Malic},
  \citenamefont {Winzer}, \citenamefont {Bobkin},\ and\ \citenamefont
  {Knorr}}]{Malic:2011}%
  \BibitemOpen
  \bibfield  {author} {\bibinfo {author} {\bibfnamefont {E.}~\bibnamefont
  {Malic}}, \bibinfo {author} {\bibfnamefont {T.}~\bibnamefont {Winzer}},
  \bibinfo {author} {\bibfnamefont {E.}~\bibnamefont {Bobkin}}, \ and\ \bibinfo
  {author} {\bibfnamefont {A.}~\bibnamefont {Knorr}},\ }\href {\doibase
  10.1103/PhysRevB.84.205406} {\bibfield  {journal} {\bibinfo  {journal} {Phys.
  Rev. B}\ }\textbf {\bibinfo {volume} {84}},\ \bibinfo {pages} {205406}
  (\bibinfo {year} {2011})}\BibitemShut {NoStop}%
\bibitem [{\citenamefont {Rohwer}\ \emph {et~al.}(2011)\citenamefont {Rohwer},
  \citenamefont {Hellmann}, \citenamefont {Wiesenmayer}, \citenamefont {Sohrt},
  \citenamefont {Stange}, \citenamefont {Slomski}, \citenamefont {Carr},
  \citenamefont {Liu}, \citenamefont {Avila}, \citenamefont {Kallane},
  \citenamefont {Mathias}, \citenamefont {Kipp}, \citenamefont {Rossnagel},\
  and\ \citenamefont {Bauer}}]{Rohwer:2011}%
  \BibitemOpen
  \bibfield  {author} {\bibinfo {author} {\bibfnamefont {T.}~\bibnamefont
  {Rohwer}}, \bibinfo {author} {\bibfnamefont {S.}~\bibnamefont {Hellmann}},
  \bibinfo {author} {\bibfnamefont {M.}~\bibnamefont {Wiesenmayer}}, \bibinfo
  {author} {\bibfnamefont {C.}~\bibnamefont {Sohrt}}, \bibinfo {author}
  {\bibfnamefont {A.}~\bibnamefont {Stange}}, \bibinfo {author} {\bibfnamefont
  {B.}~\bibnamefont {Slomski}}, \bibinfo {author} {\bibfnamefont
  {A.}~\bibnamefont {Carr}}, \bibinfo {author} {\bibfnamefont {Y.}~\bibnamefont
  {Liu}}, \bibinfo {author} {\bibfnamefont {L.~M.}\ \bibnamefont {Avila}},
  \bibinfo {author} {\bibfnamefont {M.}~\bibnamefont {Kallane}}, \bibinfo
  {author} {\bibfnamefont {S.}~\bibnamefont {Mathias}}, \bibinfo {author}
  {\bibfnamefont {L.}~\bibnamefont {Kipp}}, \bibinfo {author} {\bibfnamefont
  {K.}~\bibnamefont {Rossnagel}}, \ and\ \bibinfo {author} {\bibfnamefont
  {M.}~\bibnamefont {Bauer}},\ }\href {http://dx.doi.org/10.1038/nature09829}
  {\bibfield  {journal} {\bibinfo  {journal} {Nature}\ }\textbf {\bibinfo
  {volume} {471}},\ \bibinfo {pages} {490} (\bibinfo {year}
  {2011})}\BibitemShut {NoStop}%
\bibitem [{\citenamefont {Johannsen}\ \emph {et~al.}(2013)\citenamefont
  {Johannsen}, \citenamefont {Ulstrup}, \citenamefont {Cilento}, \citenamefont
  {Crepaldi}, \citenamefont {Zacchigna}, \citenamefont {Cacho}, \citenamefont
  {Turcu}, \citenamefont {Springate}, \citenamefont {Fromm}, \citenamefont
  {Raidel}, \citenamefont {Seyller}, \citenamefont {Parmigiani}, \citenamefont
  {Grioni},\ and\ \citenamefont {Hofmann}}]{johannsen:2013}%
  \BibitemOpen
  \bibfield  {author} {\bibinfo {author} {\bibfnamefont {J.~C.}\ \bibnamefont
  {Johannsen}}, \bibinfo {author} {\bibfnamefont {S.}~\bibnamefont {Ulstrup}},
  \bibinfo {author} {\bibfnamefont {F.}~\bibnamefont {Cilento}}, \bibinfo
  {author} {\bibfnamefont {A.}~\bibnamefont {Crepaldi}}, \bibinfo {author}
  {\bibfnamefont {M.}~\bibnamefont {Zacchigna}}, \bibinfo {author}
  {\bibfnamefont {C.}~\bibnamefont {Cacho}}, \bibinfo {author} {\bibfnamefont
  {I.~C.~E.}\ \bibnamefont {Turcu}}, \bibinfo {author} {\bibfnamefont
  {E.}~\bibnamefont {Springate}}, \bibinfo {author} {\bibfnamefont
  {F.}~\bibnamefont {Fromm}}, \bibinfo {author} {\bibfnamefont
  {C.}~\bibnamefont {Raidel}}, \bibinfo {author} {\bibfnamefont
  {T.}~\bibnamefont {Seyller}}, \bibinfo {author} {\bibfnamefont
  {F.}~\bibnamefont {Parmigiani}}, \bibinfo {author} {\bibfnamefont
  {M.}~\bibnamefont {Grioni}}, \ and\ \bibinfo {author} {\bibfnamefont
  {P.}~\bibnamefont {Hofmann}},\ }\href {\doibase
  10.1103/PhysRevLett.111.027403} {\bibfield  {journal} {\bibinfo  {journal}
  {Physical Review Letters}\ }\textbf {\bibinfo {volume} {111}},\ \bibinfo
  {pages} {027403} (\bibinfo {year} {2013})}\BibitemShut {NoStop}%
\bibitem [{\citenamefont {{Grubi\v{s}i\'c \v{C}abo}}\ \emph
  {et~al.}(2015)\citenamefont {{Grubi\v{s}i\'c \v{C}abo}}, \citenamefont
  {Miwa}, \citenamefont {Gr{\o}nborg}, \citenamefont {Riley}, \citenamefont
  {Johannsen}, \citenamefont {Cacho}, \citenamefont {Alexander}, \citenamefont
  {Chapman}, \citenamefont {Springate}, \citenamefont {Grioni}, \citenamefont
  {Lauritsen}, \citenamefont {King}, \citenamefont {Hofmann},\ and\
  \citenamefont {Ulstrup}}]{Antonija-Grubisic-Cabo:2015aa}%
  \BibitemOpen
  \bibfield  {author} {\bibinfo {author} {\bibfnamefont {A.}~\bibnamefont
  {{Grubi\v{s}i\'c \v{C}abo}}}, \bibinfo {author} {\bibfnamefont {J.~A.}\
  \bibnamefont {Miwa}}, \bibinfo {author} {\bibfnamefont {S.~S.}\ \bibnamefont
  {Gr{\o}nborg}}, \bibinfo {author} {\bibfnamefont {J.~M.}\ \bibnamefont
  {Riley}}, \bibinfo {author} {\bibfnamefont {J.~C.}\ \bibnamefont
  {Johannsen}}, \bibinfo {author} {\bibfnamefont {C.}~\bibnamefont {Cacho}},
  \bibinfo {author} {\bibfnamefont {O.}~\bibnamefont {Alexander}}, \bibinfo
  {author} {\bibfnamefont {R.~T.}\ \bibnamefont {Chapman}}, \bibinfo {author}
  {\bibfnamefont {E.}~\bibnamefont {Springate}}, \bibinfo {author}
  {\bibfnamefont {M.}~\bibnamefont {Grioni}}, \bibinfo {author} {\bibfnamefont
  {J.~V.}\ \bibnamefont {Lauritsen}}, \bibinfo {author} {\bibfnamefont
  {P.~D.~C.}\ \bibnamefont {King}}, \bibinfo {author} {\bibfnamefont
  {P.}~\bibnamefont {Hofmann}}, \ and\ \bibinfo {author} {\bibfnamefont
  {S.}~\bibnamefont {Ulstrup}},\ }\href@noop {} {\bibfield  {journal} {\bibinfo
   {journal} {Nano Lett.}\ }\textbf {\bibinfo {volume} {15}},\ \bibinfo {pages}
  {5883} (\bibinfo {year} {2015})}\BibitemShut {NoStop}%
\bibitem [{\citenamefont {Bertoni}\ \emph {et~al.}(2016)\citenamefont
  {Bertoni}, \citenamefont {Nicholson}, \citenamefont {Waldecker},
  \citenamefont {H{\"u}bener}, \citenamefont {Monney}, \citenamefont
  {Giovannini}, \citenamefont {Puppin}, \citenamefont {Hoesch}, \citenamefont
  {Springate}, \citenamefont {Chapman}, \citenamefont {Cacho}, \citenamefont
  {Wolf}, \citenamefont {Rubio},\ and\ \citenamefont
  {Ernstorfer}}]{Bertoni:2016}%
  \BibitemOpen
  \bibfield  {author} {\bibinfo {author} {\bibfnamefont {R.}~\bibnamefont
  {Bertoni}}, \bibinfo {author} {\bibfnamefont {C.~W.}\ \bibnamefont
  {Nicholson}}, \bibinfo {author} {\bibfnamefont {L.}~\bibnamefont
  {Waldecker}}, \bibinfo {author} {\bibfnamefont {H.}~\bibnamefont
  {H{\"u}bener}}, \bibinfo {author} {\bibfnamefont {C.}~\bibnamefont {Monney}},
  \bibinfo {author} {\bibfnamefont {U.~D.}\ \bibnamefont {Giovannini}},
  \bibinfo {author} {\bibfnamefont {M.}~\bibnamefont {Puppin}}, \bibinfo
  {author} {\bibfnamefont {M.}~\bibnamefont {Hoesch}}, \bibinfo {author}
  {\bibfnamefont {E.}~\bibnamefont {Springate}}, \bibinfo {author}
  {\bibfnamefont {R.~T.}\ \bibnamefont {Chapman}}, \bibinfo {author}
  {\bibfnamefont {C.}~\bibnamefont {Cacho}}, \bibinfo {author} {\bibfnamefont
  {M.}~\bibnamefont {Wolf}}, \bibinfo {author} {\bibfnamefont {A.}~\bibnamefont
  {Rubio}}, \ and\ \bibinfo {author} {\bibfnamefont {R.}~\bibnamefont
  {Ernstorfer}},\ }\href@noop {} {\bibfield  {journal} {\bibinfo  {journal}
  {Phys. Rev. Lett.}\ }\textbf {\bibinfo {volume} {117}},\ \bibinfo {pages}
  {277201} (\bibinfo {year} {2016})}\BibitemShut {NoStop}%
\bibitem [{\citenamefont {Aeschlimann}\ \emph {et~al.}(2017)\citenamefont
  {Aeschlimann}, \citenamefont {Krause}, \citenamefont {Ch\'avez-Cervantes},
  \citenamefont {Bromberger}, \citenamefont {Jago}, \citenamefont
  {Mali\ifmmode~\acute{c}\else \'{c}\fi{}}, \citenamefont {Al-Temimy},
  \citenamefont {Coletti}, \citenamefont {Cavalleri},\ and\ \citenamefont
  {Gierz}}]{Aeschlimann:2017}%
  \BibitemOpen
  \bibfield  {author} {\bibinfo {author} {\bibfnamefont {S.}~\bibnamefont
  {Aeschlimann}}, \bibinfo {author} {\bibfnamefont {R.}~\bibnamefont {Krause}},
  \bibinfo {author} {\bibfnamefont {M.}~\bibnamefont {Ch\'avez-Cervantes}},
  \bibinfo {author} {\bibfnamefont {H.}~\bibnamefont {Bromberger}}, \bibinfo
  {author} {\bibfnamefont {R.}~\bibnamefont {Jago}}, \bibinfo {author}
  {\bibfnamefont {E.}~\bibnamefont {Mali\ifmmode~\acute{c}\else \'{c}\fi{}}},
  \bibinfo {author} {\bibfnamefont {A.}~\bibnamefont {Al-Temimy}}, \bibinfo
  {author} {\bibfnamefont {C.}~\bibnamefont {Coletti}}, \bibinfo {author}
  {\bibfnamefont {A.}~\bibnamefont {Cavalleri}}, \ and\ \bibinfo {author}
  {\bibfnamefont {I.}~\bibnamefont {Gierz}},\ }\href {\doibase
  10.1103/PhysRevB.96.020301} {\bibfield  {journal} {\bibinfo  {journal} {Phys.
  Rev. B}\ }\textbf {\bibinfo {volume} {96}},\ \bibinfo {pages} {020301}
  (\bibinfo {year} {2017})}\BibitemShut {NoStop}%
\bibitem [{\citenamefont {Ulstrup}\ \emph {et~al.}(2017)\citenamefont
  {Ulstrup}, \citenamefont {{A. Grubi\v{s}i\'c \v{C}abo}}, \citenamefont
  {Biswas}, \citenamefont {Riley}, \citenamefont {Dendzik}, \citenamefont
  {Sanders}, \citenamefont {Bianchi}, \citenamefont {Cacho}, \citenamefont
  {Matselyukh}, \citenamefont {Chapman}, \citenamefont {Springate},
  \citenamefont {King}, \citenamefont {Miwa},\ and\ \citenamefont
  {Hofmann}}]{Ulstrup2017}%
  \BibitemOpen
  \bibfield  {author} {\bibinfo {author} {\bibfnamefont {S.}~\bibnamefont
  {Ulstrup}}, \bibinfo {author} {\bibnamefont {{A. Grubi\v{s}i\'c \v{C}abo}}},
  \bibinfo {author} {\bibfnamefont {D.}~\bibnamefont {Biswas}}, \bibinfo
  {author} {\bibfnamefont {J.~M.}\ \bibnamefont {Riley}}, \bibinfo {author}
  {\bibfnamefont {M.}~\bibnamefont {Dendzik}}, \bibinfo {author} {\bibfnamefont
  {C.~E.}\ \bibnamefont {Sanders}}, \bibinfo {author} {\bibfnamefont
  {M.}~\bibnamefont {Bianchi}}, \bibinfo {author} {\bibfnamefont
  {C.}~\bibnamefont {Cacho}}, \bibinfo {author} {\bibfnamefont
  {D.}~\bibnamefont {Matselyukh}}, \bibinfo {author} {\bibfnamefont {R.~T.}\
  \bibnamefont {Chapman}}, \bibinfo {author} {\bibfnamefont {E.}~\bibnamefont
  {Springate}}, \bibinfo {author} {\bibfnamefont {P.~D.~C.}\ \bibnamefont
  {King}}, \bibinfo {author} {\bibfnamefont {J.~A.}\ \bibnamefont {Miwa}}, \
  and\ \bibinfo {author} {\bibfnamefont {P.}~\bibnamefont {Hofmann}},\ }\href
  {\doibase 10.1103/PhysRevB.95.041405} {\bibfield  {journal} {\bibinfo
  {journal} {Phys. Rev. B}\ }\textbf {\bibinfo {volume} {95}},\ \bibinfo
  {pages} {041405} (\bibinfo {year} {2017})}\BibitemShut {NoStop}%
\bibitem [{\citenamefont {{Beyer}}\ \emph {et~al.}(2019)\citenamefont
  {{Beyer}}, \citenamefont {{Rohde}}, \citenamefont {{Grubi{\v{s}}i{\'c}
  {\v{C}}abo}}, \citenamefont {{Stange}}, \citenamefont {{Jacobsen}},
  \citenamefont {{Bignardi}}, \citenamefont {{Lizzit}}, \citenamefont
  {{Lacovig}}, \citenamefont {{Sanders}}, \citenamefont {{Lizzit}},
  \citenamefont {{Rossnagel}}, \citenamefont {{Hofmann}},\ and\ \citenamefont
  {{Bauer}}}]{Beyer:2019aa}%
  \BibitemOpen
  \bibfield  {author} {\bibinfo {author} {\bibfnamefont {H.}~\bibnamefont
  {{Beyer}}}, \bibinfo {author} {\bibfnamefont {G.}~\bibnamefont {{Rohde}}},
  \bibinfo {author} {\bibfnamefont {A.}~\bibnamefont {{Grubi{\v{s}}i{\'c}
  {\v{C}}abo}}}, \bibinfo {author} {\bibfnamefont {A.}~\bibnamefont
  {{Stange}}}, \bibinfo {author} {\bibfnamefont {T.}~\bibnamefont
  {{Jacobsen}}}, \bibinfo {author} {\bibfnamefont {L.}~\bibnamefont
  {{Bignardi}}}, \bibinfo {author} {\bibfnamefont {D.}~\bibnamefont
  {{Lizzit}}}, \bibinfo {author} {\bibfnamefont {P.}~\bibnamefont {{Lacovig}}},
  \bibinfo {author} {\bibfnamefont {C.~E.}\ \bibnamefont {{Sanders}}}, \bibinfo
  {author} {\bibfnamefont {S.}~\bibnamefont {{Lizzit}}}, \bibinfo {author}
  {\bibfnamefont {K.}~\bibnamefont {{Rossnagel}}}, \bibinfo {author}
  {\bibfnamefont {P.}~\bibnamefont {{Hofmann}}}, \ and\ \bibinfo {author}
  {\bibfnamefont {M.}~\bibnamefont {{Bauer}}},\ }\href@noop {} {\bibfield
  {journal} {\bibinfo  {journal} {arXiv e-prints}\ ,\ \bibinfo {eid}
  {arXiv:1907.10553}} (\bibinfo {year} {2019})},\ \Eprint
  {http://arxiv.org/abs/1907.10553} {arXiv:1907.10553 [cond-mat.mtrl-sci]}
  \BibitemShut {NoStop}%
\bibitem [{\citenamefont {Dendzik}\ \emph {et~al.}(2017)\citenamefont
  {Dendzik}, \citenamefont {Bruix}, \citenamefont {Michiardi}, \citenamefont
  {Ngankeu}, \citenamefont {Bianchi}, \citenamefont {Miwa}, \citenamefont
  {Hammer}, \citenamefont {Hofmann},\ and\ \citenamefont
  {Sanders}}]{Dendzik2017}%
  \BibitemOpen
  \bibfield  {author} {\bibinfo {author} {\bibfnamefont {M.}~\bibnamefont
  {Dendzik}}, \bibinfo {author} {\bibfnamefont {A.}~\bibnamefont {Bruix}},
  \bibinfo {author} {\bibfnamefont {M.}~\bibnamefont {Michiardi}}, \bibinfo
  {author} {\bibfnamefont {A.~S.}\ \bibnamefont {Ngankeu}}, \bibinfo {author}
  {\bibfnamefont {M.}~\bibnamefont {Bianchi}}, \bibinfo {author} {\bibfnamefont
  {J.~A.}\ \bibnamefont {Miwa}}, \bibinfo {author} {\bibfnamefont
  {B.}~\bibnamefont {Hammer}}, \bibinfo {author} {\bibfnamefont
  {P.}~\bibnamefont {Hofmann}}, \ and\ \bibinfo {author} {\bibfnamefont
  {C.~E.}\ \bibnamefont {Sanders}},\ }\href {\doibase
  10.1103/PhysRevB.96.235440} {\bibfield  {journal} {\bibinfo  {journal} {Phys.
  Rev. B}\ }\textbf {\bibinfo {volume} {96}},\ \bibinfo {pages} {235440}
  (\bibinfo {year} {2017})}\BibitemShut {NoStop}%
\bibitem [{\citenamefont {Baraldi}\ \emph {et~al.}(2003)\citenamefont
  {Baraldi}, \citenamefont {Comelli}, \citenamefont {Lizzit}, \citenamefont
  {Kiskinova},\ and\ \citenamefont {Paolucci}}]{Baraldi2003}%
  \BibitemOpen
  \bibfield  {author} {\bibinfo {author} {\bibfnamefont {A.}~\bibnamefont
  {Baraldi}}, \bibinfo {author} {\bibfnamefont {G.}~\bibnamefont {Comelli}},
  \bibinfo {author} {\bibfnamefont {S.}~\bibnamefont {Lizzit}}, \bibinfo
  {author} {\bibfnamefont {M.}~\bibnamefont {Kiskinova}}, \ and\ \bibinfo
  {author} {\bibfnamefont {G.}~\bibnamefont {Paolucci}},\ }\href {\doibase
  https://doi.org/10.1016/S0167-5729(03)00013-X} {\bibfield  {journal}
  {\bibinfo  {journal} {Surface Science Reports}\ }\textbf {\bibinfo {volume}
  {49}},\ \bibinfo {pages} {169 } (\bibinfo {year} {2003})}\BibitemShut
  {NoStop}%
\bibitem [{SMA()}]{SMAT}%
  \BibitemOpen
  \href@noop {} {\enquote {\bibinfo {title} {{See Supplemental Material for
  details of sample growth and X-ray photoelectron diffraction data on the
  distribution of BL MoS$_2$ domain orientations, for high resolution ARPES
  measurements, for TR-ARPES experimental details and for full details on the
  Hamiltonian and tight binding parameters applied in the model.}}}\
  }\BibitemShut {NoStop}%
\bibitem [{\citenamefont {He}\ \emph {et~al.}(2014)\citenamefont {He},
  \citenamefont {Hummer},\ and\ \citenamefont {Franchini}}]{He:2014}%
  \BibitemOpen
  \bibfield  {author} {\bibinfo {author} {\bibfnamefont {J.}~\bibnamefont
  {He}}, \bibinfo {author} {\bibfnamefont {K.}~\bibnamefont {Hummer}}, \ and\
  \bibinfo {author} {\bibfnamefont {C.}~\bibnamefont {Franchini}},\ }\href
  {\doibase 10.1103/PhysRevB.89.075409} {\bibfield  {journal} {\bibinfo
  {journal} {Phys. Rev. B}\ }\textbf {\bibinfo {volume} {89}},\ \bibinfo
  {pages} {075409} (\bibinfo {year} {2014})}\BibitemShut {NoStop}%
\bibitem [{\citenamefont {Hoffmann}\ \emph {et~al.}(2004)\citenamefont
  {Hoffmann}, \citenamefont {S{\o}ndergaard}, \citenamefont {Schultz},
  \citenamefont {Li},\ and\ \citenamefont {Hofmann}}]{Hoffmann:2004aa}%
  \BibitemOpen
  \bibfield  {author} {\bibinfo {author} {\bibfnamefont {S.~V.}\ \bibnamefont
  {Hoffmann}}, \bibinfo {author} {\bibfnamefont {C.}~\bibnamefont
  {S{\o}ndergaard}}, \bibinfo {author} {\bibfnamefont {C.}~\bibnamefont
  {Schultz}}, \bibinfo {author} {\bibfnamefont {Z.}~\bibnamefont {Li}}, \ and\
  \bibinfo {author} {\bibfnamefont {P.}~\bibnamefont {Hofmann}},\ }\href@noop
  {} {\bibfield  {journal} {\bibinfo  {journal} {Nucl. Instr. and Meth. Phys.
  Res. A}\ }\textbf {\bibinfo {volume} {523}},\ \bibinfo {pages} {441}
  (\bibinfo {year} {2004})}\BibitemShut {NoStop}%
\bibitem [{\citenamefont {Ulstrup}\ \emph {et~al.}(2015)\citenamefont
  {Ulstrup}, \citenamefont {Johannsen}, \citenamefont {Crepaldi}, \citenamefont
  {Cilento}, \citenamefont {Zacchigna}, \citenamefont {Cacho}, \citenamefont
  {Chapman}, \citenamefont {Springate}, \citenamefont {Fromm}, \citenamefont
  {Raidel}, \citenamefont {Seyller}, \citenamefont {Parmigiani}, \citenamefont
  {Grioni},\ and\ \citenamefont {Hofmann}}]{Ulstrup:2015bb}%
  \BibitemOpen
  \bibfield  {author} {\bibinfo {author} {\bibfnamefont {S.}~\bibnamefont
  {Ulstrup}}, \bibinfo {author} {\bibfnamefont {J.~C.}\ \bibnamefont
  {Johannsen}}, \bibinfo {author} {\bibfnamefont {A.}~\bibnamefont {Crepaldi}},
  \bibinfo {author} {\bibfnamefont {F.}~\bibnamefont {Cilento}}, \bibinfo
  {author} {\bibfnamefont {M.}~\bibnamefont {Zacchigna}}, \bibinfo {author}
  {\bibfnamefont {C.}~\bibnamefont {Cacho}}, \bibinfo {author} {\bibfnamefont
  {R.~T.}\ \bibnamefont {Chapman}}, \bibinfo {author} {\bibfnamefont
  {E.}~\bibnamefont {Springate}}, \bibinfo {author} {\bibfnamefont
  {F.}~\bibnamefont {Fromm}}, \bibinfo {author} {\bibfnamefont
  {C.}~\bibnamefont {Raidel}}, \bibinfo {author} {\bibfnamefont
  {T.}~\bibnamefont {Seyller}}, \bibinfo {author} {\bibfnamefont
  {F.}~\bibnamefont {Parmigiani}}, \bibinfo {author} {\bibfnamefont
  {M.}~\bibnamefont {Grioni}}, \ and\ \bibinfo {author} {\bibfnamefont
  {P.}~\bibnamefont {Hofmann}},\ }\href {\doibase
  10.1088/0953-8984/27/16/164206} {\bibfield  {journal} {\bibinfo  {journal}
  {Journal of Physics: Condensed Matter}\ }\textbf {\bibinfo {volume} {27}},\
  \bibinfo {pages} {164206} (\bibinfo {year} {2015})}\BibitemShut {NoStop}%
\bibitem [{\citenamefont {Rostami}\ \emph {et~al.}(2019)\citenamefont
  {Rostami}, \citenamefont {Volckaert}, \citenamefont {Lanata}, \citenamefont
  {Mahatha}, \citenamefont {Sanders}, \citenamefont {Bianchi}, \citenamefont
  {Lizzit}, \citenamefont {Bignardi}, \citenamefont {Lizzit}, \citenamefont
  {Miwa}, \citenamefont {Balatsky}, \citenamefont {Hofmann},\ and\
  \citenamefont {Ulstrup}}]{ourprb}%
  \BibitemOpen
  \bibfield  {author} {\bibinfo {author} {\bibfnamefont {H.}~\bibnamefont
  {Rostami}}, \bibinfo {author} {\bibfnamefont {K.}~\bibnamefont {Volckaert}},
  \bibinfo {author} {\bibfnamefont {N.}~\bibnamefont {Lanata}}, \bibinfo
  {author} {\bibfnamefont {S.~K.}\ \bibnamefont {Mahatha}}, \bibinfo {author}
  {\bibfnamefont {C.~E.}\ \bibnamefont {Sanders}}, \bibinfo {author}
  {\bibfnamefont {M.}~\bibnamefont {Bianchi}}, \bibinfo {author} {\bibfnamefont
  {D.}~\bibnamefont {Lizzit}}, \bibinfo {author} {\bibfnamefont
  {L.}~\bibnamefont {Bignardi}}, \bibinfo {author} {\bibfnamefont
  {S.}~\bibnamefont {Lizzit}}, \bibinfo {author} {\bibfnamefont {J.~A.}\
  \bibnamefont {Miwa}}, \bibinfo {author} {\bibfnamefont {A.~V.}\ \bibnamefont
  {Balatsky}}, \bibinfo {author} {\bibfnamefont {P.}~\bibnamefont {Hofmann}}, \
  and\ \bibinfo {author} {\bibfnamefont {S.}~\bibnamefont {Ulstrup}},\
  }\href@noop {} {\bibfield  {journal} {\bibinfo  {journal} {"Layer and orbital
  interference effects in photoemission from transition metal dichalcogenides",
  simultaneously posted on arXiv}\ } (\bibinfo {year} {2019})}\BibitemShut
  {NoStop}%
\bibitem [{\citenamefont {Moser}(2017)}]{Moser:2017}%
  \BibitemOpen
  \bibfield  {author} {\bibinfo {author} {\bibfnamefont {S.}~\bibnamefont
  {Moser}},\ }\href {\doibase https://doi.org/10.1016/j.elspec.2016.11.007}
  {\bibfield  {journal} {\bibinfo  {journal} {Journal of Electron Spectroscopy
  and Related Phenomena}\ }\textbf {\bibinfo {volume} {214}},\ \bibinfo {pages}
  {29 } (\bibinfo {year} {2017})}\BibitemShut {NoStop}%
\bibitem [{\citenamefont {Korm\'anyos}\ \emph {et~al.}(2013)\citenamefont
  {Korm\'anyos}, \citenamefont {Z\'olyomi}, \citenamefont {Drummond},
  \citenamefont {Rakyta}, \citenamefont {Burkard},\ and\ \citenamefont
  {Fal'ko}}]{Kormanyos:2013}%
  \BibitemOpen
  \bibfield  {author} {\bibinfo {author} {\bibfnamefont {A.}~\bibnamefont
  {Korm\'anyos}}, \bibinfo {author} {\bibfnamefont {V.}~\bibnamefont
  {Z\'olyomi}}, \bibinfo {author} {\bibfnamefont {N.~D.}\ \bibnamefont
  {Drummond}}, \bibinfo {author} {\bibfnamefont {P.}~\bibnamefont {Rakyta}},
  \bibinfo {author} {\bibfnamefont {G.}~\bibnamefont {Burkard}}, \ and\
  \bibinfo {author} {\bibfnamefont {V.~I.}\ \bibnamefont {Fal'ko}},\ }\href
  {\doibase 10.1103/PhysRevB.88.045416} {\bibfield  {journal} {\bibinfo
  {journal} {Phys. Rev. B}\ }\textbf {\bibinfo {volume} {88}},\ \bibinfo
  {pages} {045416} (\bibinfo {year} {2013})}\BibitemShut {NoStop}%
\bibitem [{\citenamefont {Rostami}\ \emph {et~al.}(2013)\citenamefont
  {Rostami}, \citenamefont {Moghaddam},\ and\ \citenamefont
  {Asgari}}]{Rostami:2013}%
  \BibitemOpen
  \bibfield  {author} {\bibinfo {author} {\bibfnamefont {H.}~\bibnamefont
  {Rostami}}, \bibinfo {author} {\bibfnamefont {A.~G.}\ \bibnamefont
  {Moghaddam}}, \ and\ \bibinfo {author} {\bibfnamefont {R.}~\bibnamefont
  {Asgari}},\ }\href {\doibase 10.1103/PhysRevB.88.085440} {\bibfield
  {journal} {\bibinfo  {journal} {Phys. Rev. B}\ }\textbf {\bibinfo {volume}
  {88}},\ \bibinfo {pages} {085440} (\bibinfo {year} {2013})}\BibitemShut
  {NoStop}%
\bibitem [{\citenamefont {Rostami}\ \emph {et~al.}(2016)\citenamefont
  {Rostami}, \citenamefont {Asgari},\ and\ \citenamefont
  {Guinea}}]{Rostami:2016}%
  \BibitemOpen
  \bibfield  {author} {\bibinfo {author} {\bibfnamefont {H.}~\bibnamefont
  {Rostami}}, \bibinfo {author} {\bibfnamefont {R.}~\bibnamefont {Asgari}}, \
  and\ \bibinfo {author} {\bibfnamefont {F.}~\bibnamefont {Guinea}},\ }\href
  {\doibase 10.1088/0953-8984/28/49/495001} {\bibfield  {journal} {\bibinfo
  {journal} {Journal of Physics: Condensed Matter}\ }\textbf {\bibinfo {volume}
  {28}},\ \bibinfo {pages} {495001} (\bibinfo {year} {2016})}\BibitemShut
  {NoStop}%
\bibitem [{\citenamefont {Gong}\ \emph {et~al.}(2013)\citenamefont {Gong},
  \citenamefont {Liu}, \citenamefont {Yu}, \citenamefont {Xiao}, \citenamefont
  {Cui}, \citenamefont {Xu},\ and\ \citenamefont {Yao}}]{Gong:2013}%
  \BibitemOpen
  \bibfield  {author} {\bibinfo {author} {\bibfnamefont {Z.}~\bibnamefont
  {Gong}}, \bibinfo {author} {\bibfnamefont {G.-B.}\ \bibnamefont {Liu}},
  \bibinfo {author} {\bibfnamefont {H.}~\bibnamefont {Yu}}, \bibinfo {author}
  {\bibfnamefont {D.}~\bibnamefont {Xiao}}, \bibinfo {author} {\bibfnamefont
  {X.}~\bibnamefont {Cui}}, \bibinfo {author} {\bibfnamefont {X.}~\bibnamefont
  {Xu}}, \ and\ \bibinfo {author} {\bibfnamefont {W.}~\bibnamefont {Yao}},\
  }\href {http://dx.doi.org/10.1038/ncomms3053} {\bibfield  {journal} {\bibinfo
   {journal} {Nature Communications}\ }\textbf {\bibinfo {volume} {4}},\
  \bibinfo {pages} {2053} (\bibinfo {year} {2013})}\BibitemShut {NoStop}%
\bibitem [{\citenamefont {Korm\'anyos}\ \emph {et~al.}(2018)\citenamefont
  {Korm\'anyos}, \citenamefont {Z\'olyomi}, \citenamefont {Fal'ko},\ and\
  \citenamefont {Burkard}}]{Kormanyos:2018}%
  \BibitemOpen
  \bibfield  {author} {\bibinfo {author} {\bibfnamefont {A.}~\bibnamefont
  {Korm\'anyos}}, \bibinfo {author} {\bibfnamefont {V.}~\bibnamefont
  {Z\'olyomi}}, \bibinfo {author} {\bibfnamefont {V.~I.}\ \bibnamefont
  {Fal'ko}}, \ and\ \bibinfo {author} {\bibfnamefont {G.}~\bibnamefont
  {Burkard}},\ }\href {\doibase 10.1103/PhysRevB.98.035408} {\bibfield
  {journal} {\bibinfo  {journal} {Phys. Rev. B}\ }\textbf {\bibinfo {volume}
  {98}},\ \bibinfo {pages} {035408} (\bibinfo {year} {2018})}\BibitemShut
  {NoStop}%
\bibitem [{\citenamefont {Haug}\ and\ \citenamefont
  {Koch}(2004)}]{haug2004quantum}%
  \BibitemOpen
  \bibfield  {author} {\bibinfo {author} {\bibfnamefont {H.}~\bibnamefont
  {Haug}}\ and\ \bibinfo {author} {\bibfnamefont {S.}~\bibnamefont {Koch}},\
  }\href {https://books.google.se/books?id=lJRIDQAAQBAJ} {\emph {\bibinfo
  {title} {Quantum Theory of the Optical and Electronic Properties of
  Semiconductors: Fourth Edition}}}\ (\bibinfo  {publisher} {World Scientific
  Publishing Company},\ \bibinfo {year} {2004})\BibitemShut {NoStop}%
\bibitem [{\citenamefont {Riley}\ \emph {et~al.}(2014)\citenamefont {Riley},
  \citenamefont {Mazzola}, \citenamefont {Dendzik}, \citenamefont {Michiardi},
  \citenamefont {Takayama}, \citenamefont {Bawden}, \citenamefont {Granerod},
  \citenamefont {Leandersson}, \citenamefont {Balasubramanian}, \citenamefont
  {Hoesch}, \citenamefont {Kim}, \citenamefont {Takagi}, \citenamefont
  {Meevasana}, \citenamefont {Hofmann}, \citenamefont {Bahramy}, \citenamefont
  {Wells},\ and\ \citenamefont {King}}]{Riley:2014}%
  \BibitemOpen
  \bibfield  {author} {\bibinfo {author} {\bibfnamefont {J.~M.}\ \bibnamefont
  {Riley}}, \bibinfo {author} {\bibfnamefont {F.}~\bibnamefont {Mazzola}},
  \bibinfo {author} {\bibfnamefont {M.}~\bibnamefont {Dendzik}}, \bibinfo
  {author} {\bibfnamefont {M.}~\bibnamefont {Michiardi}}, \bibinfo {author}
  {\bibfnamefont {T.}~\bibnamefont {Takayama}}, \bibinfo {author}
  {\bibfnamefont {L.}~\bibnamefont {Bawden}}, \bibinfo {author} {\bibfnamefont
  {C.}~\bibnamefont {Granerod}}, \bibinfo {author} {\bibfnamefont
  {M.}~\bibnamefont {Leandersson}}, \bibinfo {author} {\bibfnamefont
  {T.}~\bibnamefont {Balasubramanian}}, \bibinfo {author} {\bibfnamefont
  {M.}~\bibnamefont {Hoesch}}, \bibinfo {author} {\bibfnamefont {T.~K.}\
  \bibnamefont {Kim}}, \bibinfo {author} {\bibfnamefont {H.}~\bibnamefont
  {Takagi}}, \bibinfo {author} {\bibfnamefont {W.}~\bibnamefont {Meevasana}},
  \bibinfo {author} {\bibfnamefont {P.}~\bibnamefont {Hofmann}}, \bibinfo
  {author} {\bibfnamefont {M.~S.}\ \bibnamefont {Bahramy}}, \bibinfo {author}
  {\bibfnamefont {J.~W.}\ \bibnamefont {Wells}}, \ and\ \bibinfo {author}
  {\bibfnamefont {P.~D.~C.}\ \bibnamefont {King}},\ }\href
  {http://www.nature.com/nphys/journal/v10/n11/full/nphys3105.html} {\bibfield
  {journal} {\bibinfo  {journal} {Nature Physics}\ }\textbf {\bibinfo {volume}
  {10}},\ \bibinfo {pages} {835} (\bibinfo {year} {2014})}\BibitemShut
  {NoStop}%
\end{thebibliography}

\begin{thebibliography}{13}%
\makeatletter
\providecommand \@ifxundefined [1]{%
 \@ifx{#1\undefined}
}%
\providecommand \@ifnum [1]{%
 \ifnum #1\expandafter \@firstoftwo
 \else \expandafter \@secondoftwo
 \fi
}%
\providecommand \@ifx [1]{%
 \ifx #1\expandafter \@firstoftwo
 \else \expandafter \@secondoftwo
 \fi
}%
\providecommand \natexlab [1]{#1}%
\providecommand \enquote  [1]{``#1''}%
\providecommand \bibnamefont  [1]{#1}%
\providecommand \bibfnamefont [1]{#1}%
\providecommand \citenamefont [1]{#1}%
\providecommand \href@noop [0]{\@secondoftwo}%
\providecommand \href [0]{\begingroup \@sanitize@url \@href}%
\providecommand \@href[1]{\@@startlink{#1}\@@href}%
\providecommand \@@href[1]{\endgroup#1\@@endlink}%
\providecommand \@sanitize@url [0]{\catcode `\\12\catcode `\$12\catcode
  `\&12\catcode `\#12\catcode `\^12\catcode `\_12\catcode `\%12\relax}%
\providecommand \@@startlink[1]{}%
\providecommand \@@endlink[0]{}%
\providecommand \url  [0]{\begingroup\@sanitize@url \@url }%
\providecommand \@url [1]{\endgroup\@href {#1}{\urlprefix }}%
\providecommand \urlprefix  [0]{URL }%
\providecommand \Eprint [0]{\href }%
\providecommand \doibase [0]{http://dx.doi.org/}%
\providecommand \selectlanguage [0]{\@gobble}%
\providecommand \bibinfo  [0]{\@secondoftwo}%
\providecommand \bibfield  [0]{\@secondoftwo}%
\providecommand \translation [1]{[#1]}%
\providecommand \BibitemOpen [0]{}%
\providecommand \bibitemStop [0]{}%
\providecommand \bibitemNoStop [0]{.\EOS\space}%
\providecommand \EOS [0]{\spacefactor3000\relax}%
\providecommand \BibitemShut  [1]{\csname bibitem#1\endcsname}%
\let\auto@bib@innerbib\@empty
\bibitem [{\citenamefont {Baraldi}\ \emph {et~al.}(2003)\citenamefont
  {Baraldi}, \citenamefont {Comelli}, \citenamefont {Lizzit}, \citenamefont
  {Kiskinova},\ and\ \citenamefont {Paolucci}}]{Baraldi2003_2}%
  \BibitemOpen
  \bibfield  {author} {\bibinfo {author} {\bibfnamefont {A.}~\bibnamefont
  {Baraldi}}, \bibinfo {author} {\bibfnamefont {G.}~\bibnamefont {Comelli}},
  \bibinfo {author} {\bibfnamefont {S.}~\bibnamefont {Lizzit}}, \bibinfo
  {author} {\bibfnamefont {M.}~\bibnamefont {Kiskinova}}, \ and\ \bibinfo
  {author} {\bibfnamefont {G.}~\bibnamefont {Paolucci}},\ }\href {\doibase
  https://doi.org/10.1016/S0167-5729(03)00013-X} {\bibfield  {journal}
  {\bibinfo  {journal} {Surface Science Reports}\ }\textbf {\bibinfo {volume}
  {49}},\ \bibinfo {pages} {169 } (\bibinfo {year} {2003})}\BibitemShut
  {NoStop}%
\bibitem [{\citenamefont {Baker}\ \emph {et~al.}(1999)\citenamefont {Baker},
  \citenamefont {Gilmore}, \citenamefont {Lenardi},\ and\ \citenamefont
  {Gissler}}]{Baker_1999_2}%
  \BibitemOpen
  \bibfield  {author} {\bibinfo {author} {\bibfnamefont {M.}~\bibnamefont
  {Baker}}, \bibinfo {author} {\bibfnamefont {R.}~\bibnamefont {Gilmore}},
  \bibinfo {author} {\bibfnamefont {C.}~\bibnamefont {Lenardi}}, \ and\
  \bibinfo {author} {\bibfnamefont {W.}~\bibnamefont {Gissler}},\ }\href@noop
  {} {\bibfield  {journal} {\bibinfo  {journal} {Applied Surface Science}\
  }\textbf {\bibinfo {volume} {150}},\ \bibinfo {pages} {255 } (\bibinfo {year}
  {1999})}\BibitemShut {NoStop}%
\bibitem [{\citenamefont {Garc\'{\i}a~de Abajo}\ \emph
  {et~al.}(2001)\citenamefont {Garc\'{\i}a~de Abajo}, \citenamefont
  {Van~Hove},\ and\ \citenamefont {Fadley}}]{Garcia_2001_2}%
  \BibitemOpen
  \bibfield  {author} {\bibinfo {author} {\bibfnamefont {F.~J.}\ \bibnamefont
  {Garc\'{\i}a~de Abajo}}, \bibinfo {author} {\bibfnamefont {M.~A.}\
  \bibnamefont {Van~Hove}}, \ and\ \bibinfo {author} {\bibfnamefont {C.~S.}\
  \bibnamefont {Fadley}},\ }\href@noop {} {\bibfield  {journal} {\bibinfo
  {journal} {Phys. Rev. B}\ }\textbf {\bibinfo {volume} {63}},\ \bibinfo
  {pages} {075404} (\bibinfo {year} {2001})}\BibitemShut {NoStop}%
\bibitem [{\citenamefont {Gr{\o}nborg}\ \emph {et~al.}(2015)\citenamefont
  {Gr{\o}nborg}, \citenamefont {Ulstrup}, \citenamefont {Bianchi},
  \citenamefont {Dendzik}, \citenamefont {Sanders}, \citenamefont {Lauritsen},
  \citenamefont {Hofmann},\ and\ \citenamefont {Miwa}}]{Gronborg:2015aa_2}%
  \BibitemOpen
  \bibfield  {author} {\bibinfo {author} {\bibfnamefont {S.~S.}\ \bibnamefont
  {Gr{\o}nborg}}, \bibinfo {author} {\bibfnamefont {S.}~\bibnamefont
  {Ulstrup}}, \bibinfo {author} {\bibfnamefont {M.}~\bibnamefont {Bianchi}},
  \bibinfo {author} {\bibfnamefont {M.}~\bibnamefont {Dendzik}}, \bibinfo
  {author} {\bibfnamefont {C.~E.}\ \bibnamefont {Sanders}}, \bibinfo {author}
  {\bibfnamefont {J.~V.}\ \bibnamefont {Lauritsen}}, \bibinfo {author}
  {\bibfnamefont {P.}~\bibnamefont {Hofmann}}, \ and\ \bibinfo {author}
  {\bibfnamefont {J.~A.}\ \bibnamefont {Miwa}},\ }\href@noop {} {\bibfield
  {journal} {\bibinfo  {journal} {Langmuir}\ }\textbf {\bibinfo {volume}
  {31}},\ \bibinfo {pages} {9700} (\bibinfo {year} {2015})},\ \Eprint
  {http://arxiv.org/abs/http://dx.doi.org/10.1021/acs.langmuir.5b02533}
  {http://dx.doi.org/10.1021/acs.langmuir.5b02533} \BibitemShut {NoStop}%
\bibitem [{\citenamefont {Dendzik}\ \emph {et~al.}(2015)\citenamefont
  {Dendzik}, \citenamefont {Michiardi}, \citenamefont {Sanders}, \citenamefont
  {Bianchi}, \citenamefont {Miwa}, \citenamefont {Gr\o{}nborg}, \citenamefont
  {Lauritsen}, \citenamefont {Bruix}, \citenamefont {Hammer},\ and\
  \citenamefont {Hofmann}}]{Dendzik:2015aa_2}%
  \BibitemOpen
  \bibfield  {author} {\bibinfo {author} {\bibfnamefont {M.}~\bibnamefont
  {Dendzik}}, \bibinfo {author} {\bibfnamefont {M.}~\bibnamefont {Michiardi}},
  \bibinfo {author} {\bibfnamefont {C.}~\bibnamefont {Sanders}}, \bibinfo
  {author} {\bibfnamefont {M.}~\bibnamefont {Bianchi}}, \bibinfo {author}
  {\bibfnamefont {J.~A.}\ \bibnamefont {Miwa}}, \bibinfo {author}
  {\bibfnamefont {S.~S.}\ \bibnamefont {Gr\o{}nborg}}, \bibinfo {author}
  {\bibfnamefont {J.~V.}\ \bibnamefont {Lauritsen}}, \bibinfo {author}
  {\bibfnamefont {A.}~\bibnamefont {Bruix}}, \bibinfo {author} {\bibfnamefont
  {B.}~\bibnamefont {Hammer}}, \ and\ \bibinfo {author} {\bibfnamefont
  {P.}~\bibnamefont {Hofmann}},\ }\href@noop {} {\bibfield  {journal} {\bibinfo
   {journal} {Phys. Rev. B}\ }\textbf {\bibinfo {volume} {92}},\ \bibinfo
  {pages} {245442} (\bibinfo {year} {2015})}\BibitemShut {NoStop}%
\bibitem [{\citenamefont {Bana}\ \emph {et~al.}(2018)\citenamefont {Bana},
  \citenamefont {Travaglia}, \citenamefont {Bignardi}, \citenamefont {Lacovig},
  \citenamefont {Sanders}, \citenamefont {Dendzik}, \citenamefont {Michiardi},
  \citenamefont {Bianchi}, \citenamefont {Lizzit}, \citenamefont {Presel},
  \citenamefont {Angelis}, \citenamefont {Apostol}, \citenamefont {Das},
  \citenamefont {Fujii}, \citenamefont {Vobornik}, \citenamefont {Larciprete},
  \citenamefont {Baraldi}, \citenamefont {Hofmann},\ and\ \citenamefont
  {Lizzit}}]{Bana2018_2}%
  \BibitemOpen
  \bibfield  {author} {\bibinfo {author} {\bibfnamefont {H.}~\bibnamefont
  {Bana}}, \bibinfo {author} {\bibfnamefont {E.}~\bibnamefont {Travaglia}},
  \bibinfo {author} {\bibfnamefont {L.}~\bibnamefont {Bignardi}}, \bibinfo
  {author} {\bibfnamefont {P.}~\bibnamefont {Lacovig}}, \bibinfo {author}
  {\bibfnamefont {C.~E.}\ \bibnamefont {Sanders}}, \bibinfo {author}
  {\bibfnamefont {M.}~\bibnamefont {Dendzik}}, \bibinfo {author} {\bibfnamefont
  {M.}~\bibnamefont {Michiardi}}, \bibinfo {author} {\bibfnamefont
  {M.}~\bibnamefont {Bianchi}}, \bibinfo {author} {\bibfnamefont
  {D.}~\bibnamefont {Lizzit}}, \bibinfo {author} {\bibfnamefont
  {F.}~\bibnamefont {Presel}}, \bibinfo {author} {\bibfnamefont {D.~D.}\
  \bibnamefont {Angelis}}, \bibinfo {author} {\bibfnamefont {N.}~\bibnamefont
  {Apostol}}, \bibinfo {author} {\bibfnamefont {P.~K.}\ \bibnamefont {Das}},
  \bibinfo {author} {\bibfnamefont {J.}~\bibnamefont {Fujii}}, \bibinfo
  {author} {\bibfnamefont {I.}~\bibnamefont {Vobornik}}, \bibinfo {author}
  {\bibfnamefont {R.}~\bibnamefont {Larciprete}}, \bibinfo {author}
  {\bibfnamefont {A.}~\bibnamefont {Baraldi}}, \bibinfo {author} {\bibfnamefont
  {P.}~\bibnamefont {Hofmann}}, \ and\ \bibinfo {author} {\bibfnamefont
  {S.}~\bibnamefont {Lizzit}},\ }\href
  {http://stacks.iop.org/2053-1583/5/i=3/a=035012} {\bibfield  {journal}
  {\bibinfo  {journal} {2D Materials}\ }\textbf {\bibinfo {volume} {5}},\
  \bibinfo {pages} {035012} (\bibinfo {year} {2018})}\BibitemShut {NoStop}%
\bibitem [{\citenamefont {Woodruff}(2007)}]{xpd_rfac_2}%
  \BibitemOpen
  \bibfield  {author} {\bibinfo {author} {\bibfnamefont {D.}~\bibnamefont
  {Woodruff}},\ }\href@noop {} {\bibfield  {journal} {\bibinfo  {journal}
  {Surf. Sci. Rep.}\ }\textbf {\bibinfo {volume} {62}},\ \bibinfo {pages} {1}
  (\bibinfo {year} {2007})}\BibitemShut {NoStop}%
\bibitem [{\citenamefont {Pendry}(1980)}]{Pendry_1980_2}%
  \BibitemOpen
  \bibfield  {author} {\bibinfo {author} {\bibfnamefont {J.~B.}\ \bibnamefont
  {Pendry}},\ }\href {\doibase 10.1088/0022-3719/13/5/024} {\bibfield
  {journal} {\bibinfo  {journal} {Journal of Physics C: Solid State Physics}\
  }\textbf {\bibinfo {volume} {13}},\ \bibinfo {pages} {937} (\bibinfo {year}
  {1980})}\BibitemShut {NoStop}%
\bibitem [{\citenamefont {Hoffmann}\ \emph {et~al.}(2004)\citenamefont
  {Hoffmann}, \citenamefont {S{\o}ndergaard}, \citenamefont {Schultz},
  \citenamefont {Li},\ and\ \citenamefont {Hofmann}}]{Hoffmann:2004aa_2}%
  \BibitemOpen
  \bibfield  {author} {\bibinfo {author} {\bibfnamefont {S.~V.}\ \bibnamefont
  {Hoffmann}}, \bibinfo {author} {\bibfnamefont {C.}~\bibnamefont
  {S{\o}ndergaard}}, \bibinfo {author} {\bibfnamefont {C.}~\bibnamefont
  {Schultz}}, \bibinfo {author} {\bibfnamefont {Z.}~\bibnamefont {Li}}, \ and\
  \bibinfo {author} {\bibfnamefont {P.}~\bibnamefont {Hofmann}},\ }\href@noop
  {} {\bibfield  {journal} {\bibinfo  {journal} {Nucl. Instr. and Meth. Phys.
  Res. A}\ }\textbf {\bibinfo {volume} {523}},\ \bibinfo {pages} {441}
  (\bibinfo {year} {2004})}\BibitemShut {NoStop}%
\bibitem [{\citenamefont {He}\ \emph {et~al.}(2014)\citenamefont {He},
  \citenamefont {Hummer},\ and\ \citenamefont {Franchini}}]{He:2014_2}%
  \BibitemOpen
  \bibfield  {author} {\bibinfo {author} {\bibfnamefont {J.}~\bibnamefont
  {He}}, \bibinfo {author} {\bibfnamefont {K.}~\bibnamefont {Hummer}}, \ and\
  \bibinfo {author} {\bibfnamefont {C.}~\bibnamefont {Franchini}},\ }\href
  {\doibase 10.1103/PhysRevB.89.075409} {\bibfield  {journal} {\bibinfo
  {journal} {Phys. Rev. B}\ }\textbf {\bibinfo {volume} {89}},\ \bibinfo
  {pages} {075409} (\bibinfo {year} {2014})}\BibitemShut {NoStop}%
\bibitem [{\citenamefont {Rostami}\ \emph {et~al.}(2019)\citenamefont
  {Rostami}, \citenamefont {Volckaert}, \citenamefont {Lanata}, \citenamefont
  {Mahatha}, \citenamefont {Sanders}, \citenamefont {Bianchi}, \citenamefont
  {Lizzit}, \citenamefont {Bignardi}, \citenamefont {Lizzit}, \citenamefont
  {Miwa}, \citenamefont {Balatsky}, \citenamefont {Hofmann},\ and\
  \citenamefont {Ulstrup}}]{ourprb_2}%
  \BibitemOpen
  \bibfield  {author} {\bibinfo {author} {\bibfnamefont {H.}~\bibnamefont
  {Rostami}}, \bibinfo {author} {\bibfnamefont {K.}~\bibnamefont {Volckaert}},
  \bibinfo {author} {\bibfnamefont {N.}~\bibnamefont {Lanata}}, \bibinfo
  {author} {\bibfnamefont {S.~K.}\ \bibnamefont {Mahatha}}, \bibinfo {author}
  {\bibfnamefont {C.~E.}\ \bibnamefont {Sanders}}, \bibinfo {author}
  {\bibfnamefont {M.}~\bibnamefont {Bianchi}}, \bibinfo {author} {\bibfnamefont
  {D.}~\bibnamefont {Lizzit}}, \bibinfo {author} {\bibfnamefont
  {L.}~\bibnamefont {Bignardi}}, \bibinfo {author} {\bibfnamefont
  {S.}~\bibnamefont {Lizzit}}, \bibinfo {author} {\bibfnamefont {J.~A.}\
  \bibnamefont {Miwa}}, \bibinfo {author} {\bibfnamefont {A.~V.}\ \bibnamefont
  {Balatsky}}, \bibinfo {author} {\bibfnamefont {P.}~\bibnamefont {Hofmann}}, \
  and\ \bibinfo {author} {\bibfnamefont {S.}~\bibnamefont {Ulstrup}},\
  }\href@noop {} {\bibfield  {journal} {\bibinfo  {journal} {"Layer and orbital
  interference effects in photoemission from transition metal dichalcogenides",
  simultaneously posted on arXiv}\ } (\bibinfo {year} {2019})}\BibitemShut
  {NoStop}%
\bibitem [{\citenamefont {Gong}\ \emph {et~al.}(2013)\citenamefont {Gong},
  \citenamefont {Liu}, \citenamefont {Yu}, \citenamefont {Xiao}, \citenamefont
  {Cui}, \citenamefont {Xu},\ and\ \citenamefont {Yao}}]{Gong:2013_2}%
  \BibitemOpen
  \bibfield  {author} {\bibinfo {author} {\bibfnamefont {Z.}~\bibnamefont
  {Gong}}, \bibinfo {author} {\bibfnamefont {G.-B.}\ \bibnamefont {Liu}},
  \bibinfo {author} {\bibfnamefont {H.}~\bibnamefont {Yu}}, \bibinfo {author}
  {\bibfnamefont {D.}~\bibnamefont {Xiao}}, \bibinfo {author} {\bibfnamefont
  {X.}~\bibnamefont {Cui}}, \bibinfo {author} {\bibfnamefont {X.}~\bibnamefont
  {Xu}}, \ and\ \bibinfo {author} {\bibfnamefont {W.}~\bibnamefont {Yao}},\
  }\href {http://dx.doi.org/10.1038/ncomms3053} {\bibfield  {journal} {\bibinfo
   {journal} {Nature Communications}\ }\textbf {\bibinfo {volume} {4}},\
  \bibinfo {pages} {2053} (\bibinfo {year} {2013})}\BibitemShut {NoStop}%
\bibitem [{\citenamefont {Korm\'anyos}\ \emph {et~al.}(2018)\citenamefont
  {Korm\'anyos}, \citenamefont {Z\'olyomi}, \citenamefont {Fal'ko},\ and\
  \citenamefont {Burkard}}]{Kormanyos:2018_2}%
  \BibitemOpen
  \bibfield  {author} {\bibinfo {author} {\bibfnamefont {A.}~\bibnamefont
  {Korm\'anyos}}, \bibinfo {author} {\bibfnamefont {V.}~\bibnamefont
  {Z\'olyomi}}, \bibinfo {author} {\bibfnamefont {V.~I.}\ \bibnamefont
  {Fal'ko}}, \ and\ \bibinfo {author} {\bibfnamefont {G.}~\bibnamefont
  {Burkard}},\ }\href {\doibase 10.1103/PhysRevB.98.035408} {\bibfield
  {journal} {\bibinfo  {journal} {Phys. Rev. B}\ }\textbf {\bibinfo {volume}
  {98}},\ \bibinfo {pages} {035408} (\bibinfo {year} {2018})}\BibitemShut
  {NoStop}%
\end{thebibliography}
\end{document}